\documentclass[aps,prd,reprint,superscriptaddress,longbibliography]{revtex4-1}

\usepackage[utf8]{inputenc}
\usepackage{amssymb}
\usepackage{amsmath}
\usepackage{hyperref}
\usepackage{graphicx}
\usepackage{booktabs}
\usepackage{verbatim}
\usepackage[nowatermark]{fixmetodonotes}
\usepackage[table]{xcolor}
\usepackage{diagbox} 
\usepackage{isotope}
\usepackage{braket}
\usepackage{siunitx}
\usepackage{cleveref}
\usepackage[labelformat=simple]{subcaption}

\usepackage{multirow} 
\usepackage{slashbox}
\definecolor{table_color}{rgb}{0.0,0.45,0.0}
\usepackage{float}
\usepackage{caption}
\graphicspath{{Images/}}

\begin{document}

\title{Comparing generator predictions of transverse kinematic imbalance in neutrino-argon scattering}

\author{Lars Bathe-Peters}
\email[]{lars\_bathe-peters@fas.harvard.edu}
\affiliation{Harvard University, Cambridge MA 02138, USA}
\author{Steven Gardiner}
\affiliation{Fermi National Accelerator Laboratory, Batavia IL 60510, USA}
\author{Roxanne Guenette}
\affiliation{Harvard University, Cambridge MA 02138, USA}

\date{\today}

\begin{abstract}
The largest uncertainties in estimating neutrino-nucleus interaction cross sections lie in the incomplete understanding of nuclear effects. A powerful tool to study nuclear effects is Transverse Kinematic Imbalance. This paper presents the first detailed comparison of the predictions of multiple event generators for distributions associated with Transverse Kinematic Imbalance for neutrino interactions on argon. Predictions for muon neutrinos interacting with an argon target are obtained using four standard neutrino event generation tools (GENIE, NuWro, GiBUU and NEUT). Example opportunities for discrimination between nuclear models leveraging future measurements are highlighted. The predictions shown in this paper are motivated by studying muon neutrinos from the Fermilab Booster Neutrino Beam interacting at the location of the MicroBooNE liquid argon time projection chamber, but the methods directly apply to other accelerator-based liquid argon neutrino experiments such as SBND, ICARUS and DUNE.
\end{abstract}

\maketitle

\section{Introduction}
High precision measurements by current and future neutrino experiments require a good understanding of neutrino-matter interactions. A variety of theoretically-challenging nuclear
effects need to be correctly taken into account in order to precisely calculate neutrino-nucleus interaction cross sections. Deficiencies in existing models predicting these effects represent a
leading source of systematic uncertainty for neutrino oscillation experiments \cite{Alvarez_Ruso_2018,Benhar:2013bwa,Megias:2019pol,PhysRevD.92.051302,Lu:2018stk,Munteanu:2019llq}.
Obtaining a better understanding of this physics is a main goal of current experimental efforts at accelerator neutrino energies~\cite{Dolan:2018zye,Formaggio:2013kya,PhysRevC.94.015503,Furmanski:2016wqo,Katori:2016yel,Alvarez-Ruso:2014bla}.
Experimental analyses rely upon models of neutrino scattering processes as implemented in Monte Carlo event generators such as GENIE~\cite{Andreopoulos:2009rq}, NEUT~\cite{Hayato:2009zz}, NuWro~\cite{golan2012,golan2014}, and GiBUU~\cite{Mosel_2019,buss2012}. Recent cross-section measurements on hydrocarbon performed by the T2K \cite{PhysRevD.98.032003} and MINER$\nu$A \cite{PhysRevD.101.092001,Lu:2018stk} collaborations have demonstrated the power of using a particular set of observables associated with Transverse Kinematic Imbalance (TKI) \cite{PhysRevC.94.015503} to constrain theoretical uncertainties related to neutrino-nucleus scattering. In light of the upcoming high-precision measurements of neutrino oscillations pursued by the Short-Baseline Neutrino (SBN) program~\cite{Machado2019,Acciarri2015}
and the Deep Underground Neutrino Experiment (DUNE) \cite{DUNETDRvol1}, which
will employ large liquid argon time projection chambers (LArTPCs) to detect
neutrinos, detailed measurements of TKI in neutrino-argon scattering are likely to provide valuable model constraints. These can in turn lead to improvements in neutrino event generators.

In particular, to make precision measurements of
neutrino oscillations, it is crucial to be able to reconstruct the original neutrino energy from the final-state particles of each neutrino interaction. This task requires a good understanding of the relevant neutrino interaction physics.
At low energies, the major contributions to inclusive neutrino-nucleus cross sections are quasi-elastic (QE) scattering, $n$ particle $n$ hole ($n$p-$n$h) interactions, and resonant pion production (RES)~\cite{PhysRevLett.74.4384,Formaggio:2013kya}. In this study, we consider these processes in the charged-current (CC) channel.

\begin{figure}[htb!]
	\centering
	\includegraphics[width=1.\columnwidth]{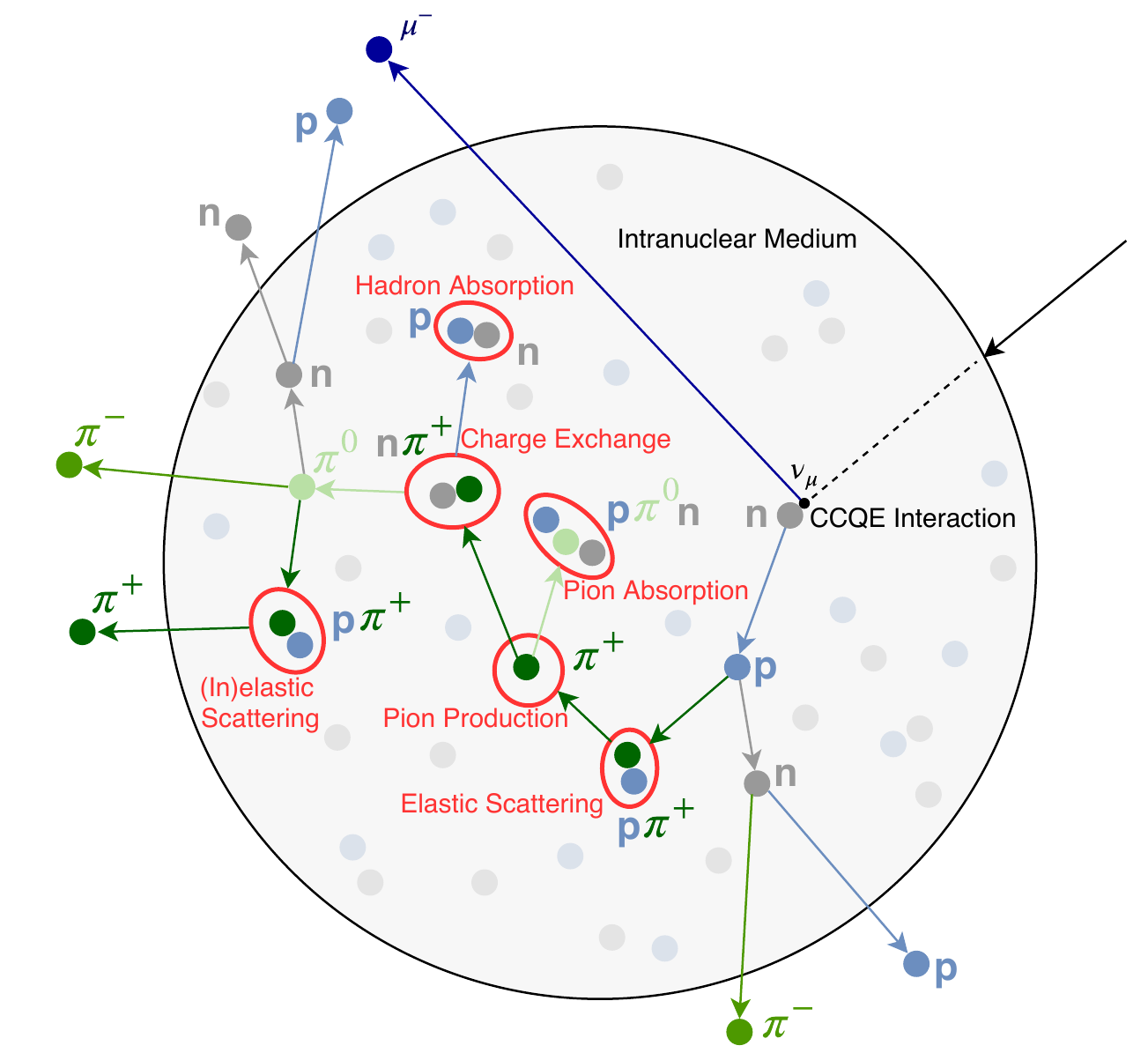}
	\caption[Illustration of Final-State Interactions]{Illustration of final-state interactions in an intranuclear cascade model approach. The incoming muon neutrino $\nu_\mu$ represented by the black arrow followed by the dashed black line interacts via a charged-current quasi-elastic interaction with a neutron in an argon nucleus.
	The outgoing muon (indigo line) and proton (light blue line) traverse the nuclear medium. The proton can undergo various types of hadronic FSIs such as (in)elastic scattering, pion production, hadron absorption and charge exchange. Pions in this figure are shown in dark green ($\pi^+$), regular green ($\pi^-$), and light green ($\pi^0$). }
	\label{FSIs}
\end{figure}

Among the most important nuclear effects in neutrino scattering are the initial-state motion of nucleons (Fermi motion), the $n$p-$n$h interactions (for $n \geq 2$) that arise due to multinucleon correlation effects, also referred to as meson exchange currents (MECs), and hadronic final-state interactions (FSIs). This paper focuses on the latter two categories of nuclear effects. The $n$p-$n$h contribution
is a topic of sustained interest in the theoretical community. Clarifying the degree to which it can be distinguished from look-alike processes~\cite{Dolan:2018sbb} in argon will be valuable moving forward.
Final-state interactions are typically described using approximate models implemented in the aforementioned neutrino interaction generators~\cite{Dytman2021}. An illustration of various possible FSIs that may occur following a primary neutrino interaction is presented in Fig.~\ref{FSIs}. 
Since FSIs can obscure the nature of the primary neutrino interaction (and thus complicate attempts at neutrino energy reconstruction), the understanding of FSIs and their implementations in modern neutrino event generators is of great
importance.

After investigating the sensitivity of the TKI variables for nuclear effects, the ultimate goal of this analysis is to find regions of phase space that allow the discrimination of neutrino interaction models with respect to a measurement of TKI variables in MicroBooNE or other experiments based on LArTPC technology. In this paper, sensitivity of TKI is investigated for an experimental setup similar to MicroBooNE to study the power of these variables and to prepare for a possible future comparison of the presented theoretical predictions with experimental data. Such comparisons are invaluable to inform how current neutrino interaction models can be modified to reduce uncertainties in neutrino measurements. In Sec.~\ref{results}, the muon and proton momentum thresholds were chosen to match a realistic signal definition for a prospective measurement of TKI in MicroBooNE.

This paper is structured in the following way: After introducing TKI in Sec.~\ref{stki_intro}, we provide an overview of the implementation of theoretical models for neutrino-nucleus scattering in several standard event generators described in Sec.~\ref{generators}. In Sec.~\ref{results}, we then present the first detailed comparisons of multiple event generator predictions of TKI distributions for neutrino-argon scattering in a detector like MicroBooNE~\cite{Acciarri2017}.

\section{Transverse Kinematic Imbalance}
\label{stki_intro}
\begin{figure}[hbt!]
	\begin{subfigure}[b]{.49\textwidth}
	\hspace{-8mm}
		\includegraphics[width=0.89\textwidth]{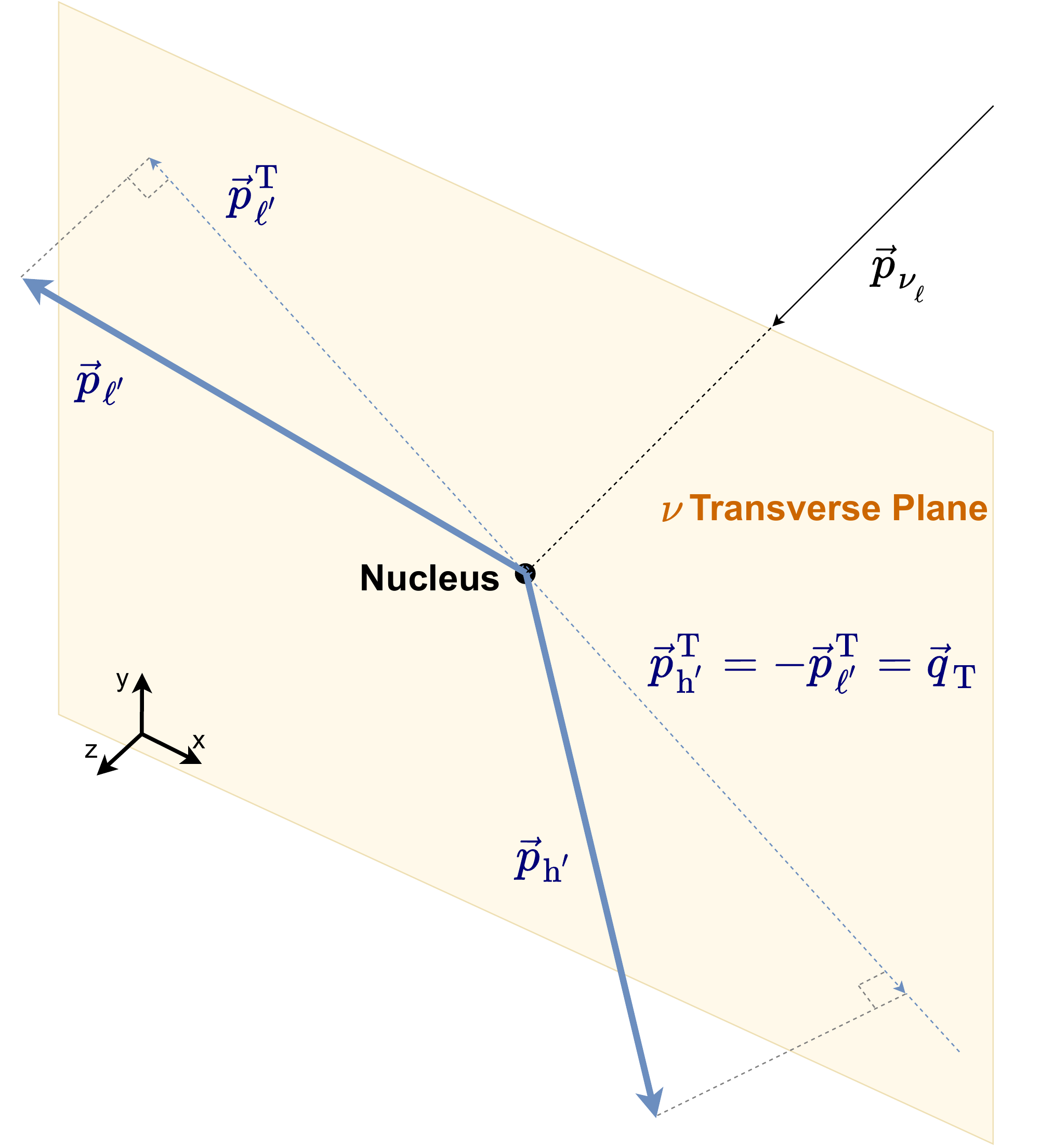}
		\caption{}
		\label{stkb}
	\end{subfigure}
	\begin{subfigure}[b]{.49\textwidth}
		\includegraphics[width=0.96\textwidth]{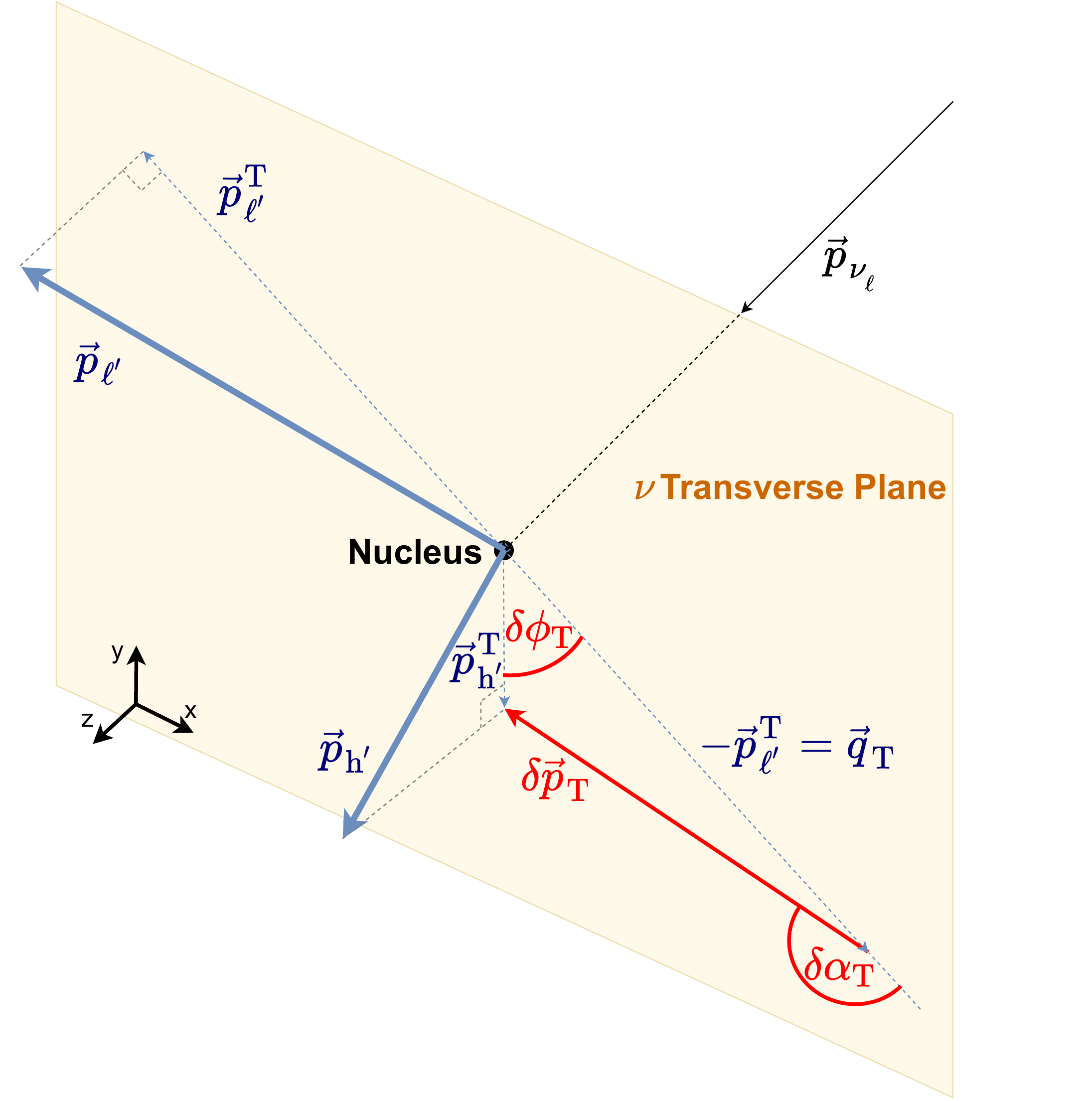}
		\caption{}
		\label{stki_stv}
	\end{subfigure}
	\caption[]{Illustration of Transverse Kinematic Imbalance (TKI) in a charged-current neutrino interaction. The three-momenta $\vec{p}_{\ell'}$ and $\vec{p}_{\text{h}'}$ of outgoing particles are projected onto the plane transverse to the incident neutrino direction. \subref{stkb} The transverse momenta of the final-state lepton $\ell'$ and hadron $\text{h}'$ are exactly balanced in the absence of nuclear effects. \subref{stki_stv} Nuclear effects result in an imbalance between the transverse final-state particle momenta. The TKI variables $\delta \vec{p}_\text{T}$, $\delta\phi_\text{T}$ and $\delta\alpha_\text{T}$ characterise this imbalance.}
	\label{stki}
\end{figure}

A charged-current interaction, where an incoming neutrino of flavor $\ell \in \{\text{e},\mu,\tau\}$ interacts with a nucleon
will result in one charged lepton $\ell'$ and one or more hadrons
in the final state. In the case of a pure CCQE neutrino 
(antineutrino)
interaction without FSIs, there can only be one proton (neutron)
and no mesons in the final state. As illustrated in
Fig.~\ref{stki}, to study TKI, the momentum vectors $\vec{p}_{\ell'}$ and $\vec{p}_{\text{h}'}$ of the charged lepton and outgoing hadron (respectively) are projected onto the plane transverse to the direction of the incoming neutrino with momentum $\vec{p}_{\nu_{_\ell}}$. This procedure defines the transverse lepton ($\vec{p}^{\hspace{.2ex} \text{T}}_{\ell'}$) and hadron ($\vec{p}_{\text{h}'}^{\hspace{.2ex} \text{T}}$) momenta. In the absence of nuclear effects these transverse vectors are equal in magnitude but oriented in the opposite direction from one another due to conservation of momentum.
The presence of nuclear effects leads to an alteration of the interaction kinematics, so that there is a deviation from $\vec{p}^{\hspace{.2ex} \text{T}}_{\ell'} = -\vec{p}_{\text{h}'}^{\hspace{.2ex} \text{T}}$ (see Fig.~\ref{stki_stv}). This Transverse Kinematic Imbalance (TKI) may be characterised by a set of three variables~\cite{PhysRevD.92.051302,PhysRevC.94.015503}, which are depicted in red in~Fig.~\ref{stki}. The mathematical definitions of these variables are as follows:
\begin{subequations} \label{STVs}
	\begin{align}
	\delta \vec{p}_\text{T} &\equiv \vec{p}_{\ell'}^{\hspace{.2ex} \text{T}} + \vec{p}_{\text{h}'}^{\hspace{.2ex} \text{T}} \label{stvp} \\
	\delta\phi_\text{T} &\equiv \text{arccos}\left(\frac{-\vec{p}_{\ell'}^{ \hspace{.2ex} \text{T}} \cdot \delta \vec{p}_\text{T}}{p_{\ell'}^{\hspace{.2ex} \text{T}} \delta p_\text{T}}\right) \label{stvphi} \\
	\delta\alpha_\text{T} &\equiv \text{arccos}\left(\frac{-\vec{p}_{\ell'}^{\hspace{.2ex} \text{T}} \cdot \vec{p}_{\text{h}'}^{\hspace{.2ex} \text{T}}}{p_{\ell'}^\text{T}p_{\text{h}'}^\text{T}}\right) \label{stva}
	\end{align}
\end{subequations}
The vector transverse momentum imbalance $\delta \vec{p}_\text{T}$ is the sum of the transverse projections of the lepton and hadron momenta. In the absence of nuclear effects, no TKI is expected, and thus  $\delta \vec{p}_\text{T} = 0$. The transverse angular imbalance $\delta\alpha_\text{T}$ gives information about the direction of the transverse momentum imbalance $\delta \vec{p}_\text{T}$ and on how much the final-state hadron is transversely
accelerated $(|\vec{p}_{\text{h}'}^{\hspace{.2ex} \text{T}}|
< |\vec{p}^{\hspace{.2ex} \text{T}}_{\ell'}| \ \text{and} \ \delta\alpha_\text{T} < 90^\circ)$ or decelerated
$(|\vec{p}_{\text{h}'}^{\hspace{.2ex} \text{T}}|
> |\vec{p}^{\hspace{.2ex} \text{T}}_{\ell'}| \ \text{and} \ \delta\alpha_\text{T} > 90^\circ)$ 
by nuclear effects causing the TKI~\cite{Lu:2018stk}. The observable $\delta\phi_\text{T}$ is the angle between $\vec{p}_{\text{h}'}^{\hspace{.2ex} \text{T}}$ and $-\vec{p}_{\ell'}^{\hspace{.2ex} \text{T}}$. Weaker nuclear effects, causing less TKI, will shift the distribution towards smaller values of $\delta\phi_\text{T}$ where $\delta\phi_\text{T} = 0$ corresponds to perfect balance between the transverse momenta.

It is possible to construct additional TKI variables in events with several hadrons $\text{h}_i$ in the final state, where the total transverse momentum imbalance is defined by
\begin{align} \label{dpt_nh}
    \delta \vec{p}_\text{T}^{\hspace{.3ex} n\text{h}} \equiv \vec{p}_{\ell'}^{\hspace{.2ex} \text{T}} + \sum_{i=1}^{n} \vec{p}_{\text{h}'_i}^{\hspace{.2ex} \text{T}}\,.
\end{align}
An interesting special case where two protons, but no other hadrons, come out of the nucleus in the final state yields the observable
\begin{align} \label{dpt_2p}
    \delta \vec{p}_\text{T}^{\hspace{.2ex} 2\text{p}} \equiv \vec{p}_{\ell'}^{\hspace{.2ex} \text{T}} + \vec{p}^{\hspace{.2ex} \text{T}}_{\text{p}_1} + \vec{p}^{\hspace{.2ex} \text{T}}_{\text{p}_2}
\end{align}
This TKI variable considers events with two protons $\text{p}_1$ and $\text{p}_2$ alongside the muon in the final state and can be useful to investigate the impact of nuclear effects in CC$2\text{p}0\pi$ $n$p-$n$h events or CCQE-like events where an additional proton escapes the nucleus due to
FSIs. The superior proton reconstruction capabilities of LArTPC neutrino detectors may enable a first measurement of TKI in two-proton events. These events, which have begun to be investigated already by MicroBooNE~\cite{uBooNE2p}, are expected to have a strong $2$p-$2$h component which is poorly constrained by other data so far. An illustration of these additional TKI variables can be seen in Fig.~\ref{STKI_add}.

Previous TKI measurements using a hydrocarbon target in the T2K~\cite{T2K:2021naz,PhysRevD.98.032003} and MINER$\nu$A~\cite{MINERvA:2020anu,PhysRevD.101.092001} experiments have already demonstrated that inclusion of the nuclear effects described above is necessary for neutrino interaction models to provide an adequate description of the experimental data. Some degree of model discrimination was also achieved, but the conclusions that may be derived therefrom for ultimate application in LArTPC-based analyses are somewhat uncertain, in part due to the need to extrapolate to the heavier argon nucleus. Future direct measurements of TKI in neutrino-argon scattering, such as those explored phenomenologically in the latter part of this paper, will thus provide a complementary input to efforts to improve neutrino interaction modeling for the worldwide precision oscillation program.

\section{Neutrino event generators}
\label{generators}
Experimental studies of accelerator neutrinos rely on detailed simulations of beam production, neutrino scattering, final-state particle transport, and the detector response in order to interpret data. Neutrino event generators implement models of neutrino cross sections in a manner which is convenient for use in experimental simulation workflows. A challenge faced by developers of neutrino event generators in this energy regime is the lack of a comprehensive theoretical framework that can predict all observables for all interaction modes from first principles. Thus, the standard approach at present is for event generators to adopt a factorization scheme in which model components are joined together to fully simulate the scattering process. A core assumption in most cases is the \textit{impulse approximation}: the primary neutrino interaction is taken to involve scattering on a single bound nucleon, and corrections are applied for the effects of the nuclear environment.

Simulation of a typical neutrino-nucleus scattering event first involves sampling the interaction mode and the initial kinematics (momentum, removal energy) of the struck nucleon. A local Fermi gas (LFG) description of the nuclear ground state is used in the configurations of all four standard event generators (GENIE, NuWro, GiBUU and NEUT) considered in this paper. With the initial state fully specified, outgoing kinematics for the primary final-state particles are chosen according to the differential cross section appropriate for the sampled interaction mode. The event is complete only after possible intranuclear rescattering of the outgoing hadrons is treated using an FSI model (see Fig.~\ref{FSIs}). 
While GENIE, NuWro, and NEUT simulate FSIs using variants of an intranuclear cascade approach, GiBUU employs Boltzmann-Uehling-Uhlenbeck (BUU) transport theory to describe the same physics.

In the generator comparisons that follow, differences in the treatment of nuclear effects (as opposed to the primary scattering process) are the main focus. To minimize trivial differences between the generators, several minor modifications were made to the generator configurations. The procedure needed to reproduce these changes (as well as all other calculations described in this paper) starting from unaltered installations of GENIE~3.0.6, NuWro $19.02.1$, GiBUU $2019$ and NEUT $5.4.0$ is described in Appendix~\ref{model_elements}.

Recent releases (major version 3) of the GENIE neutrino event generator include multiple ``comprehensive model configurations'' which are officially maintained~\cite{GENIEepj}. For the results shown in this paper, we used GENIE~3.0.6 with a slightly modified version of the \texttt{G18\_10b\_00\_000} configuration. In this GENIE model set, CCQE scattering is simulated using the Nieves model~\cite{Nieves_PhysRevC.83.045501}. A dipole axial form factor is used together with the BBBA07~\cite{BBBA07} parameterization of the vector form factors. Long-range nucleon-nucleon correlations are accounted for in this model via the Random Phase Approximation (RPA). The CC$n$p-$n$h model was also developed by Nieves and collaborators~\cite{Nieves:2011yp}. Resonance production and coherent pion production (COH) are treated using the model of Berger and Sehgal \cite{Berger:2008xs}. A model based on the calculation of Bodek and Yang~\cite{Bodek:2002vp} is used for deep inelastic scattering (DIS). Hadronic FSIs are simulated using an intranuclear
cascade (INC) model called $hN$~\cite{Dytman2021}.

In NuWro 19.02.1, the Llewelyn-Smith model~\cite{LLEWELLYNSMITH1972261} is used for quasielastic cross sections. Corrections for long-range correlations are included via the RPA, but the treatment of these is that of Graczyk and Sobczyk~\cite{NuWroRPA} rather than Nieves. The QE vector and axial form factor models are set to BBBA05 and a dipole model, respectively. The $n$p-$n$h interaction cross section is described by the Nieves model. Resonance production is handled according to the Adler-Rarita-Schwinger model~\cite{ADLER1968189} instead of the Rein-Sehgal model used by the other generators. The Rein-Sehgal model is used for COH interactions. DIS scattering is based on a quark-parton model described in Ref.~\cite{golan2014}, where high-energy resonances are averaged
by structure functions. In the INC model implemented in NuWro, particles traversing through the nuclear medium are considered as classical objects moving along straight lines between collisions. The implementation is based on the Metropolis~\cite{PhysRev.110.185} and Oset~\cite{SALCEDO1988557} cascade models. While many details are different, it is conceptually similar to the GENIE $hN$ treatment.

In GiBUU $2019$, in addition to the LFG ground-state nuclear model with a local nuclear potential~\cite{Leitner:2008ue}, there is a Coulomb potential~\cite{Mosel_2019}. The GiBUU framework has its own QE and RES cross section models defined in \cite{Leitner:2008ue}. The quasielastic axial form factor uses the dipole parameterization, and the BBBA05 form is used for the vector form factors. In the configuration used in this paper, Nieves-like RPA corrections are applied for CCQE interactions. The $2$p-$2$h cross-section model
is based on an analysis of semi-inclusive electron scattering data~\cite{Bosted:2012qc,PhysRevC.81.055213}.
The dependence of this $2$p-$2$h cross-section model on the third component $T_3$ of the nuclear isospin~\cite{Gallmeister:2016dnq} leads to a
particularly strong contribution of this channel for argon in comparison to a hydrocarbon target. This feature is not shared by the $2$p-$2$h models in the other generators studied herein. A data-driven DIS model is used, and no COH events are simulated.
Hadronic final-state interactions are handled in GiBUU $2019$ by numerically solving the Boltzmann-Uehling-Uhlenbeck (BUU) differential equation.
This semi-classical approach reduces to an intranuclear cascade if certain simplifying approximations are introduced~\cite{Mosel_2019}.

For the NEUT~5.4.0 generator, CCQE and 2p-2h interactions are simulated using the Nieves model. The BBBA05 vector form factors are used together with a dipole parameterization for the axial-vector form factor. The Nieves model RPA corrections are applied for the CCQE cross section. The
Rein-Sehgal model is used in NEUT for the RES and COH channels.
The Bodek-Yang model is used for DIS. FSIs are simulated via a semi-classical INC model that is $hN$-like~\cite{Hayato:2009zz,HAYATO2002171}.

\section{Results}
\label{results}
To obtain the predictions for TKI differential cross sections discussed in the remainder of this paper, we created large samples of inclusive muon neutrino CC events using all four generator configurations documented in Sec.~\ref{generators} and Appendix~\ref{model_elements}.
The published estimate of the muon neutrino flux from the Booster Neutrino Beam at the location of the MicroBooNE detector~\cite{PhysRevLett.123.131801} was used as input to the simulations. In our calculations of flux-averaged differential cross sections, a Monte Carlo event is considered part of the signal if it includes one muon with a momentum of at least $\SI{0.150}{\GeV/\text{c}}$ and at least one proton with a momentum at or above $\SI{0.3}{\GeV/\text{c}}$ in the final state. These kinematic limits are imposed to mimic the thresholds that might be adopted in a future experimental analysis. In particular, the proton momentum threshold exactly matches the one achieved in two recent MicroBooNE cross-section measurements~\cite{uBooNECCNp,uBooNECCQElike}.

\begin{figure}[htbp!]
	\begin{subfigure}[b]{.48\textwidth}
		\includegraphics[width=\textwidth]{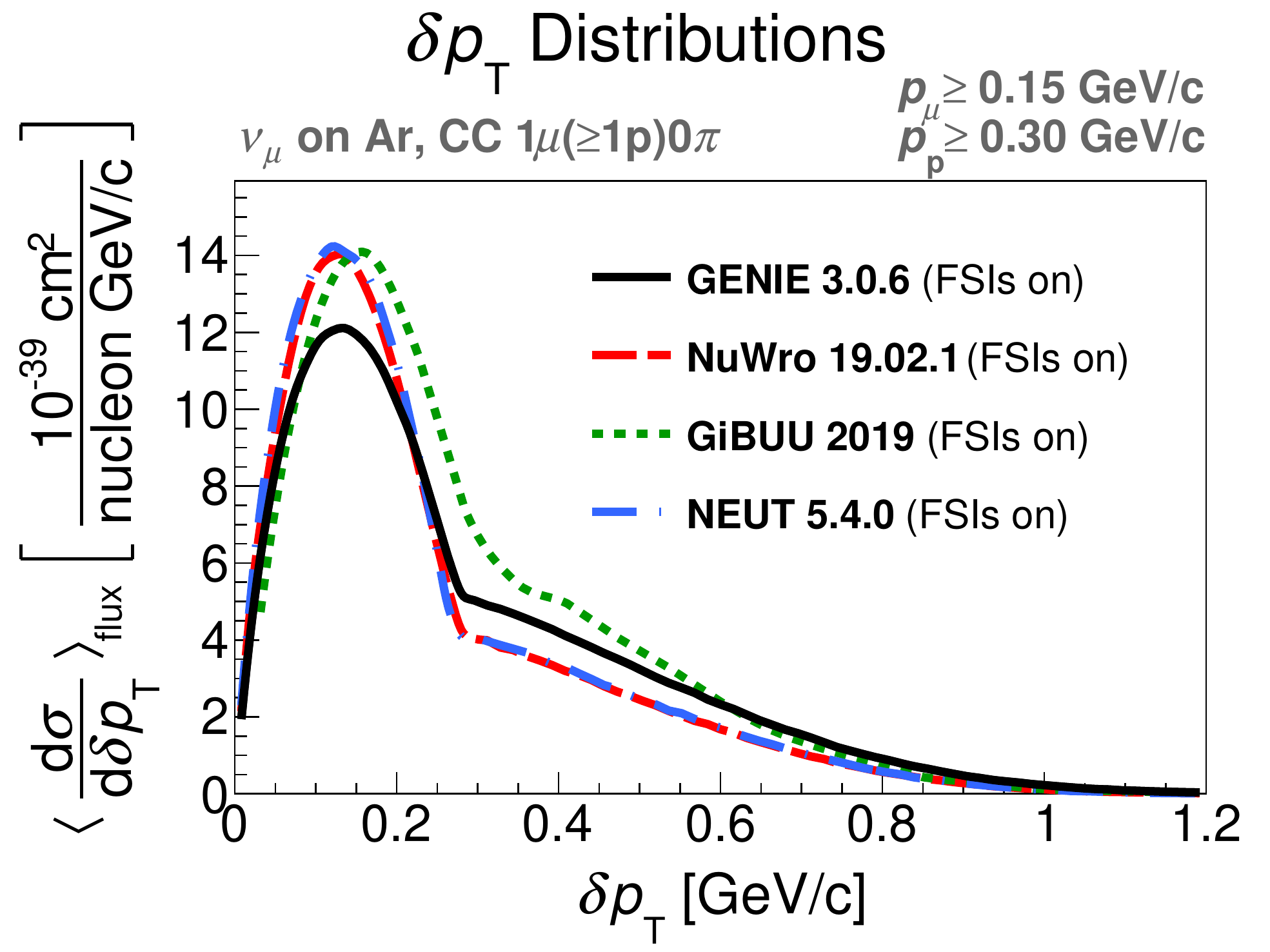}
		\vspace{-0.6cm}
		\caption{}
		\label{fsi_gencomp_dpt}
	\end{subfigure}
	\begin{subfigure}[b]{.48\textwidth}
		\includegraphics[width=\textwidth]{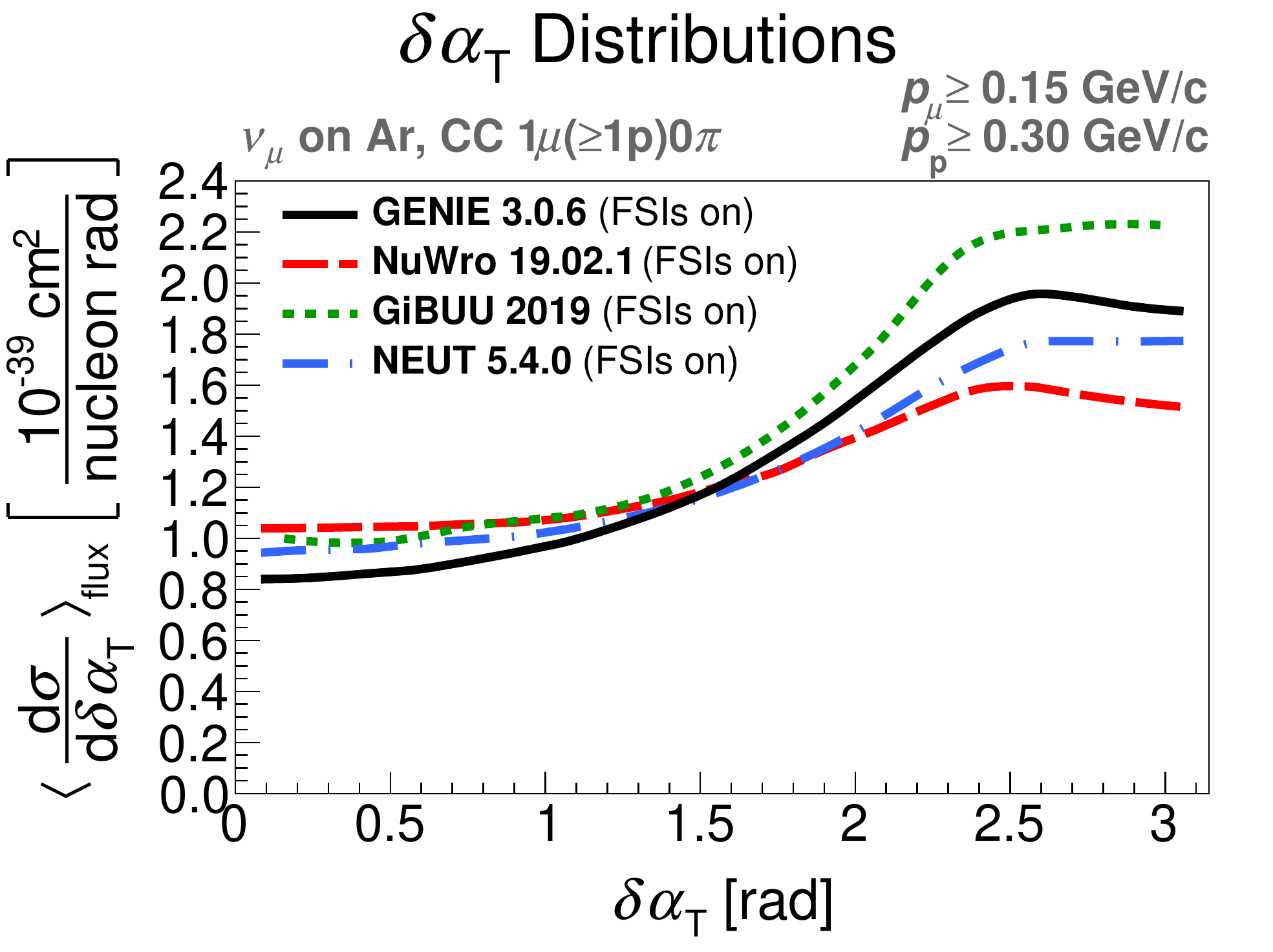}
		\vspace{-0.6cm}
		\caption{}
		\label{fsi_gencomp_dalphat}
	\end{subfigure}
	\begin{subfigure}[b]{.48\textwidth}
		\includegraphics[width=\textwidth]{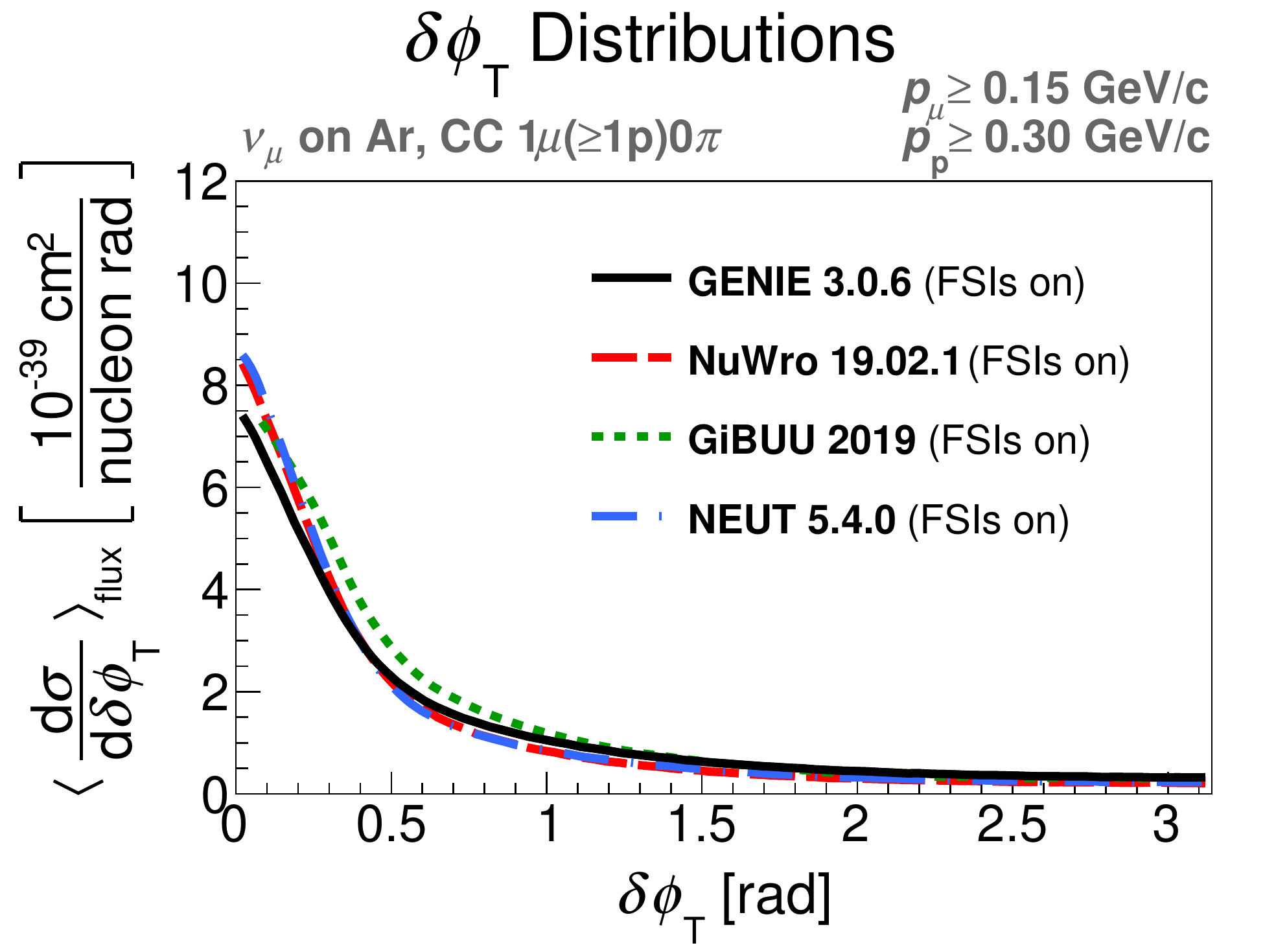}
		\vspace{-0.6cm}
		\caption{}
		\label{fsi_gencomp_dphit}
	\end{subfigure}
	\caption[Simulator Comparisons for Investigating the Impact and Discrimination of Nuclear Effects in Neutrino Interactions on Argon]{Differential cross section of muon neutrino charged-current interactions on argon as a function of three TKI variables $\delta p_\text{T}$ (\subref{fsi_gencomp_dpt}), $\delta\alpha_\text{T}$ (\subref{fsi_gencomp_dalphat}) and $\delta\phi_\text{T}$ (\subref{fsi_gencomp_dphit}) for different event generators. For all selected events, the leading proton momentum is at least $\SI{0.300}{\GeV/\text{c}}$ and the muon momentum is at least $\SI{0.150}{\GeV/\text{c}}$.}
	\label{fsi_gencomp}
\end{figure}

Figure~\ref{fsi_gencomp} shows the flux-averaged single-differential cross sections predicted by the four event generators for the three TKI variables defined in Sec.~\ref{stki_intro}.
\begin{figure}[b!]
	\begin{subfigure}[b]{.48\textwidth}
        \includegraphics[width=\textwidth]{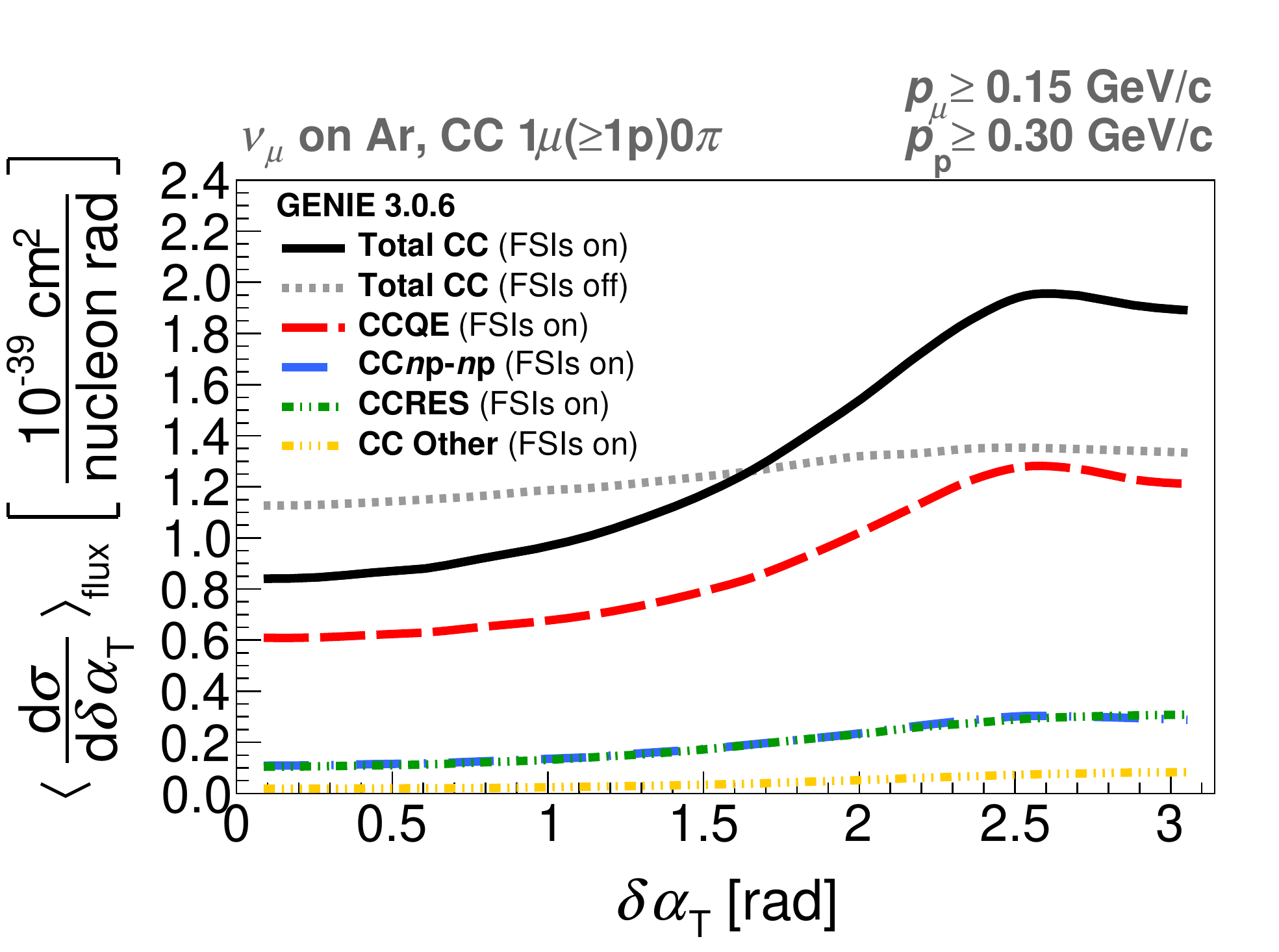}
        \vspace{-0.6cm}
        \caption{}
        \label{fsi_genie_split_dalphat}
	\end{subfigure}
        \vskip\baselineskip\vspace{-0.50cm}
	\begin{subfigure}[b]{.48\textwidth}
        \includegraphics[width=\textwidth]{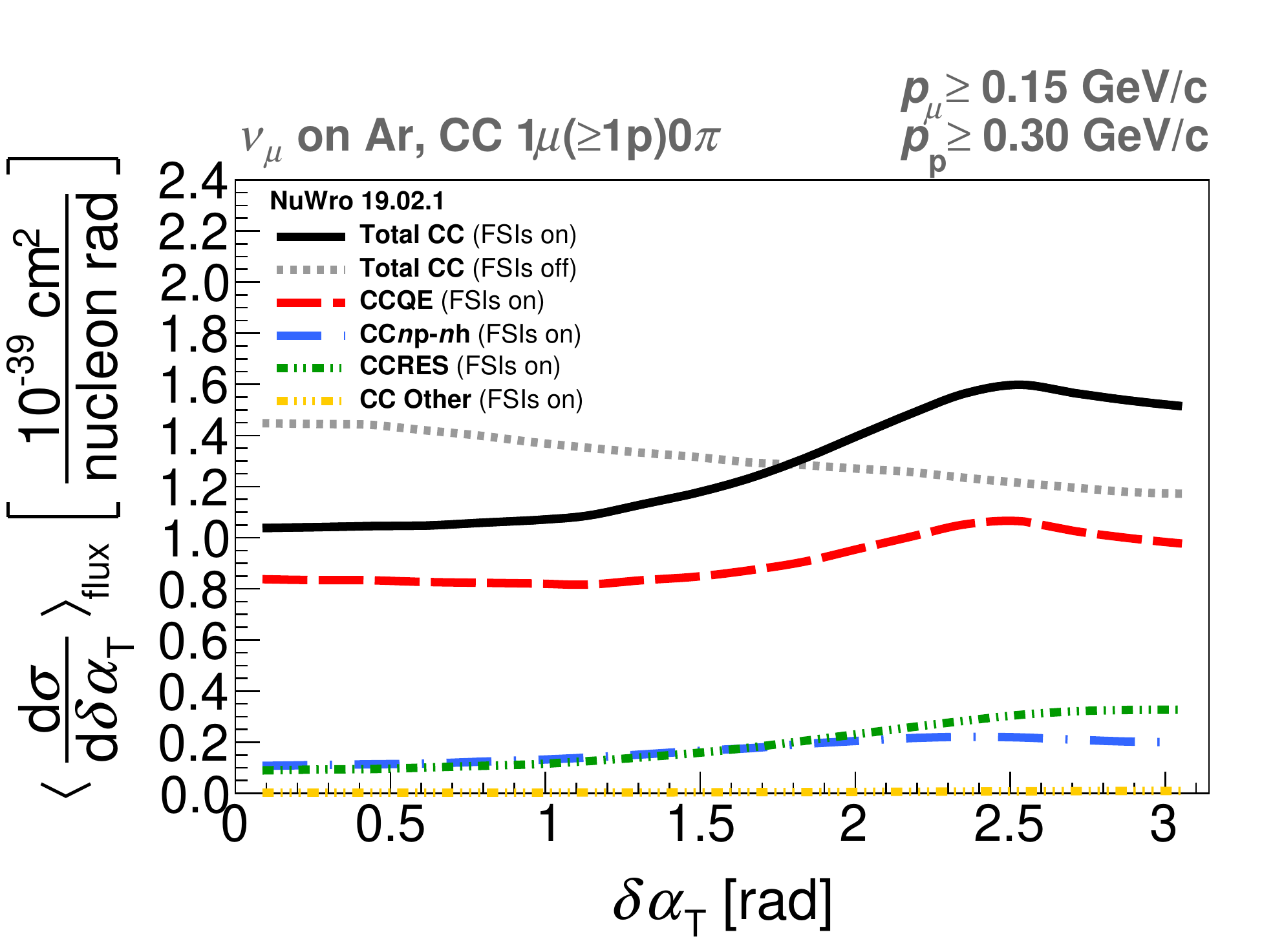}
        \vspace{-0.6cm}
        \caption{}
        \label{fsi_nuwro_split_dalphat}
	\end{subfigure}
        \vskip\baselineskip\vspace{-0.50cm}
	\begin{subfigure}[b]{.48\textwidth}
        \includegraphics[width=\textwidth]{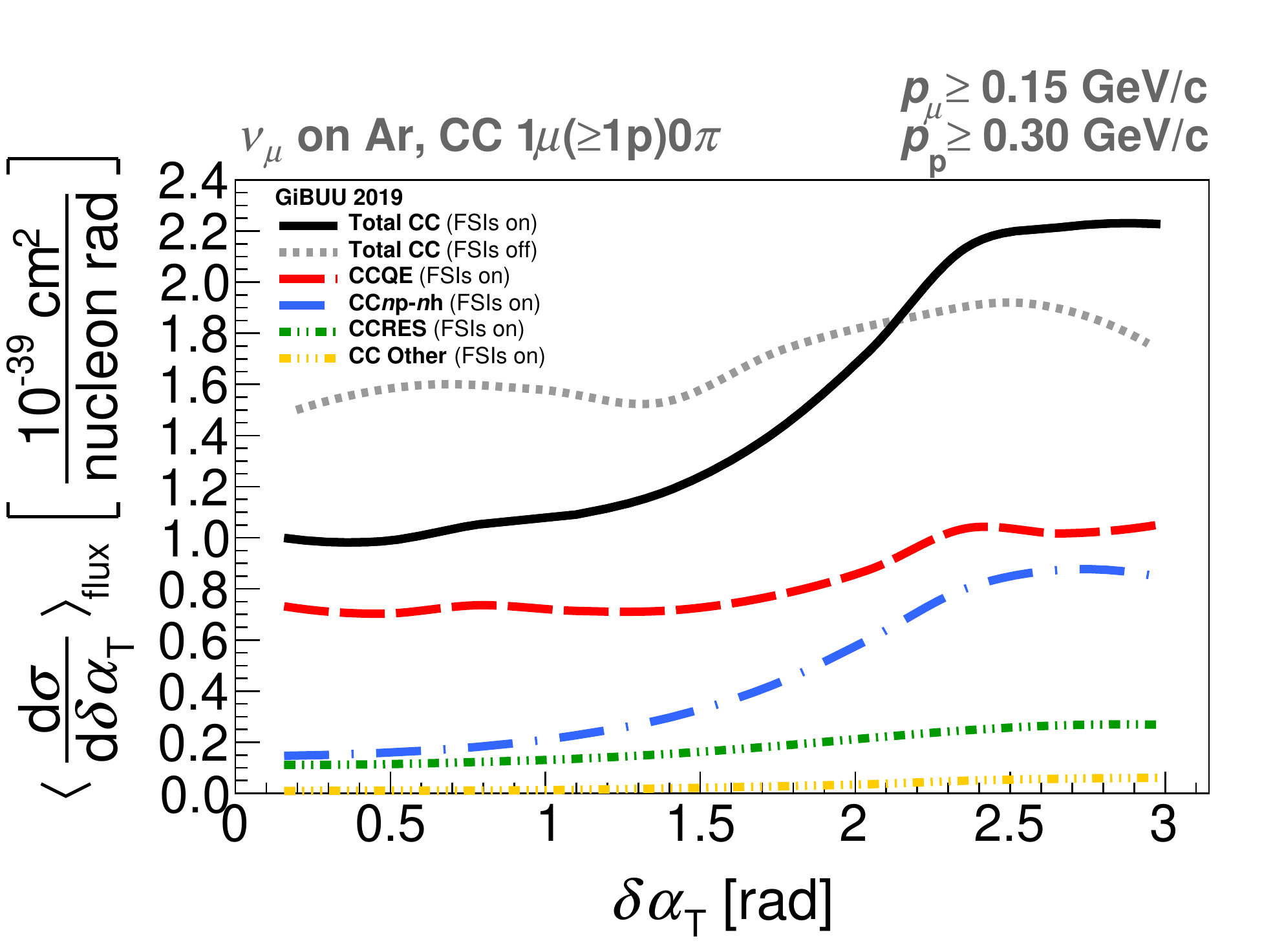}
        \vspace{-0.6cm}
        \caption{}
        \label{fsi_gibuu_split_dalphat}
	\end{subfigure}
\end{figure}
\begin{figure}
\ContinuedFloat
	\begin{subfigure}[b]{.48\textwidth}
        \includegraphics[width=\textwidth]{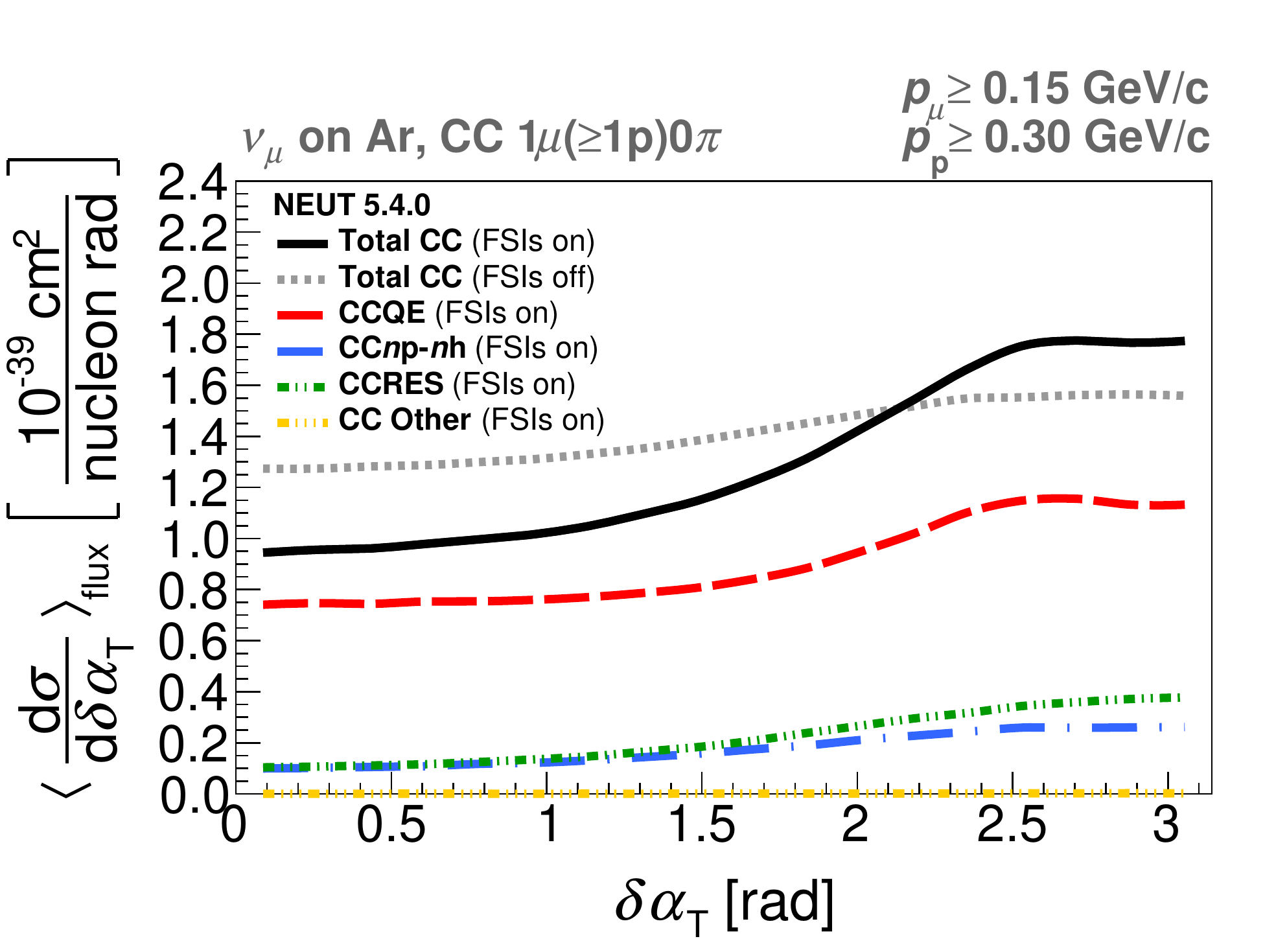}
        \vspace{-0.6cm}
        \caption{}
        \label{fsi_neut_split_dalphat}
	\end{subfigure}
	\caption[Simulator Comparisons for Investigating the Sensitivity of Nuclear Effects on the Transverse Angular Imbalance $\delta\alpha_\text{T}$ in Neutrino Interactions on Argon]{Differential cross section of muon neutrino CC interactions on argon as a function of $\delta\alpha_\text{T}$ for different true interaction channels (colored lines) produced by the four generators studied ((a) to (d)). The total cross section is also shown with (black line) and without (grey dotted line) FSIs enabled in the simulation. For all selected events, the leading proton momentum is at least $\SI{0.300}{\GeV/\text{c}}$ and the muon momentum is at least $\SI{0.150}{\GeV/\text{c}}$.}
	\label{gencomp_dalphat}
\end{figure}
While the predictions for the $\delta\phi_\text{T}$ distribution shown in Fig.~\ref{fsi_gencomp_dphit} are nearly identical for all of the models studied, there are notable discrepancies between the simulated results for the other two TKI variables. In particular, there is a large difference in the shape of the $\delta\alpha_\text{T}$ distribution shown in Fig.~\ref{fsi_gencomp_dalphat}, with GiBUU predicting the greatest asymmetry between the high- and low-$\delta\alpha_\text{T}$ regions and NuWro predicting the least. 

\subsection{The asymmetry of \texorpdfstring{$\mathbf{\textit{d}}\boldsymbol{\sigma}/\textit{d}\boldsymbol{\delta\alpha}_\text{T}$}{TEXT}}

While the potential for model discrimination in a future measurement of $d\sigma/d\delta\alpha_\text{T}$ is already apparent from the figure, a more detailed interpretation may be obtained by considering the contributions of individual reaction modes (QE, $n$p-$n$h, RES, etc.). This is done for each of the four generators in Fig.~\ref{gencomp_dalphat}. Each subfigure includes a curve for the total differential cross section (solid black) as well as partial differential cross sections calculated using only QE (dashed red), $n$p-$n$h (dash-dotted blue), and RES (dash-double-dotted green) events. A small contribution from events which do not fall into any other category (dash-triple-dotted yellow) is also shown, as well as an alternate total differential cross section for which FSIs have been entirely neglected (dotted gray). Comparisons between Figs.~\ref{fsi_genie_split_dalphat}--\ref{fsi_neut_split_dalphat} reveal a clear disagreement between the generators about the primary driver of the asymmetry of the $\delta\alpha_\text{T}$ distribution seen in Fig.~\ref{fsi_gencomp_dalphat}. For GENIE, NuWro, and NEUT, the asymmetry is largely due to CCQE events with subsequent proton FSIs. On the other hand, the results from GiBUU in Fig.~\ref{fsi_gibuu_split_dalphat} suggest that the main culprit is the contribution of $n$p-$n$h interactions. Because the shape of $d\sigma/d\delta\alpha_\text{T}$ is simultaneously sensitive to both effects, a lone measurement of this kinematic distribution would not be able to distinguish between these alternatives. However, with a sufficiently high-statistics LArTPC data set, an measurement of a multi-differential distribution could potentially show a preference for one of the two competing explanations.

Fig.~\ref{fsi_gencomp_dpt_dalphat_slices} considers one possibility of this kind: a double-differential measurement expressed as the $d\sigma/d\delta p_\text{T}$ distribution in four bins of $\delta\alpha_\text{T}$.
\begin{figure}[b!]
	\begin{subfigure}[b]{.48\textwidth}
		\includegraphics[width=\textwidth]{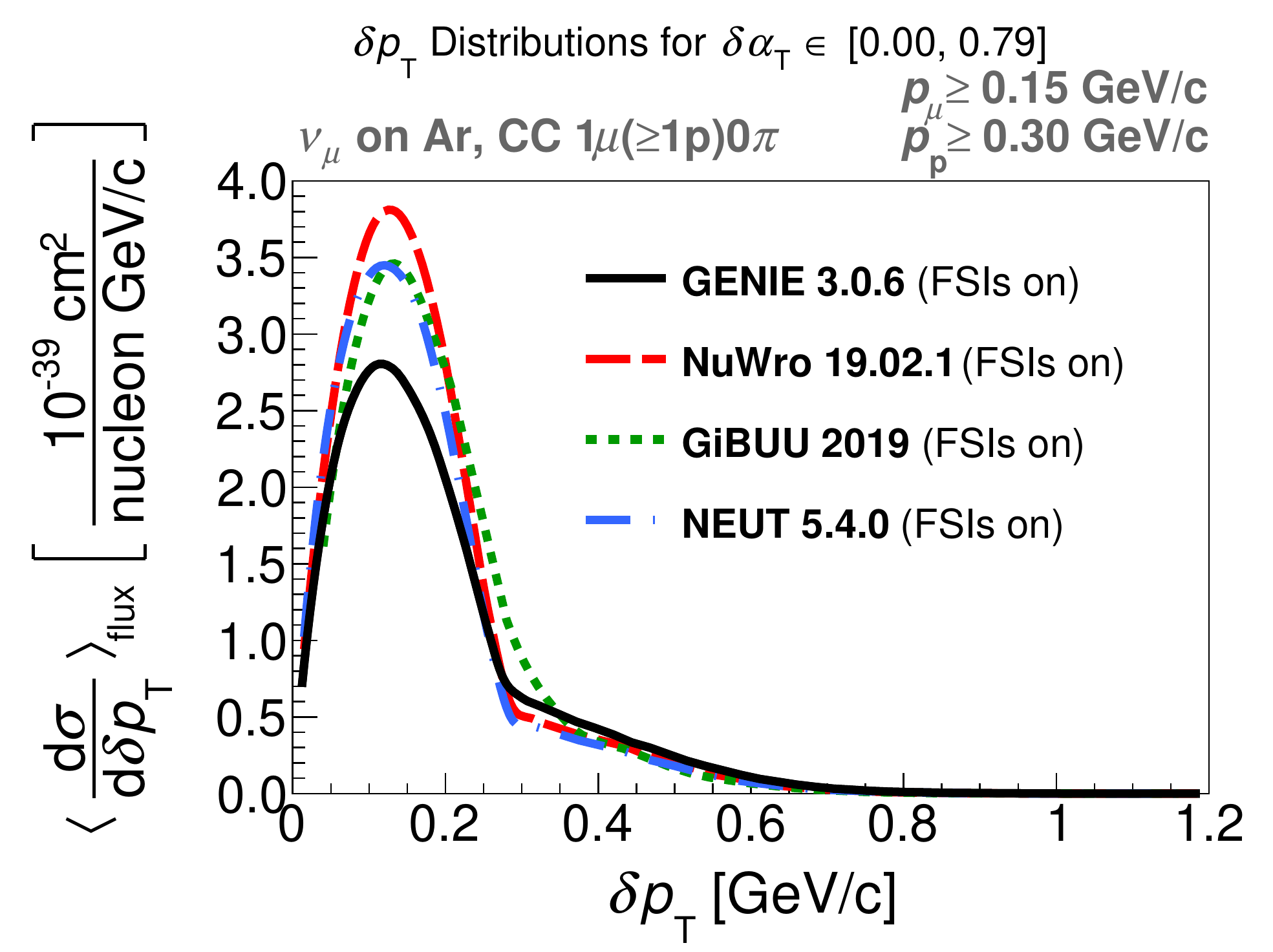}
		\vspace{-6.mm}
		\caption{}
		\label{fsi_gencomp_dpt_dalphat_1}
	\end{subfigure}
	\begin{subfigure}[b]{.48\textwidth}
		\includegraphics[width=\textwidth]{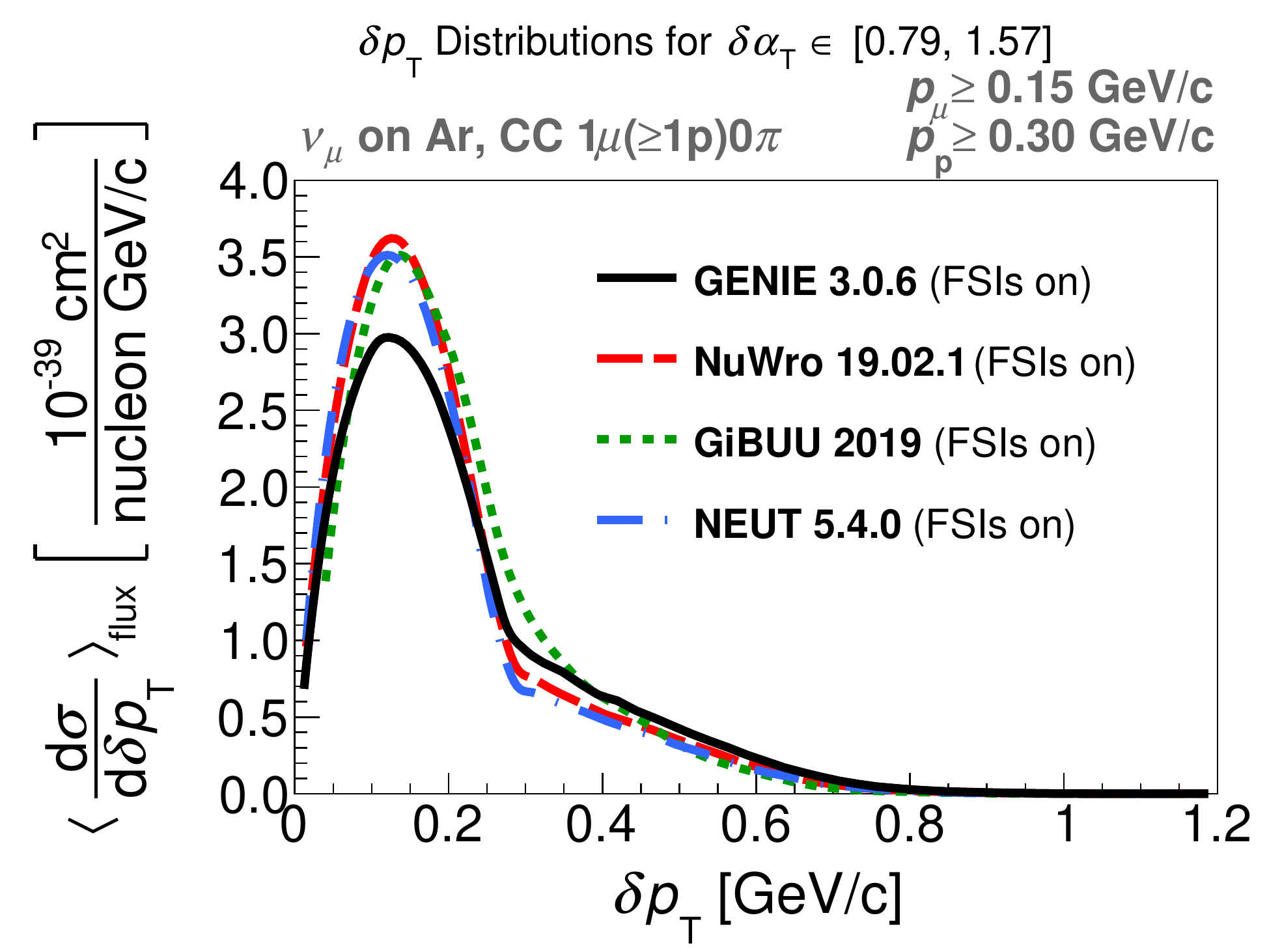}
		\vspace{-6.mm}
		\caption{}
		\label{fsi_gencomp_dpt_dalphat_2}
	\end{subfigure}
	\begin{subfigure}[b]{.48\textwidth}
		\includegraphics[width=\textwidth]{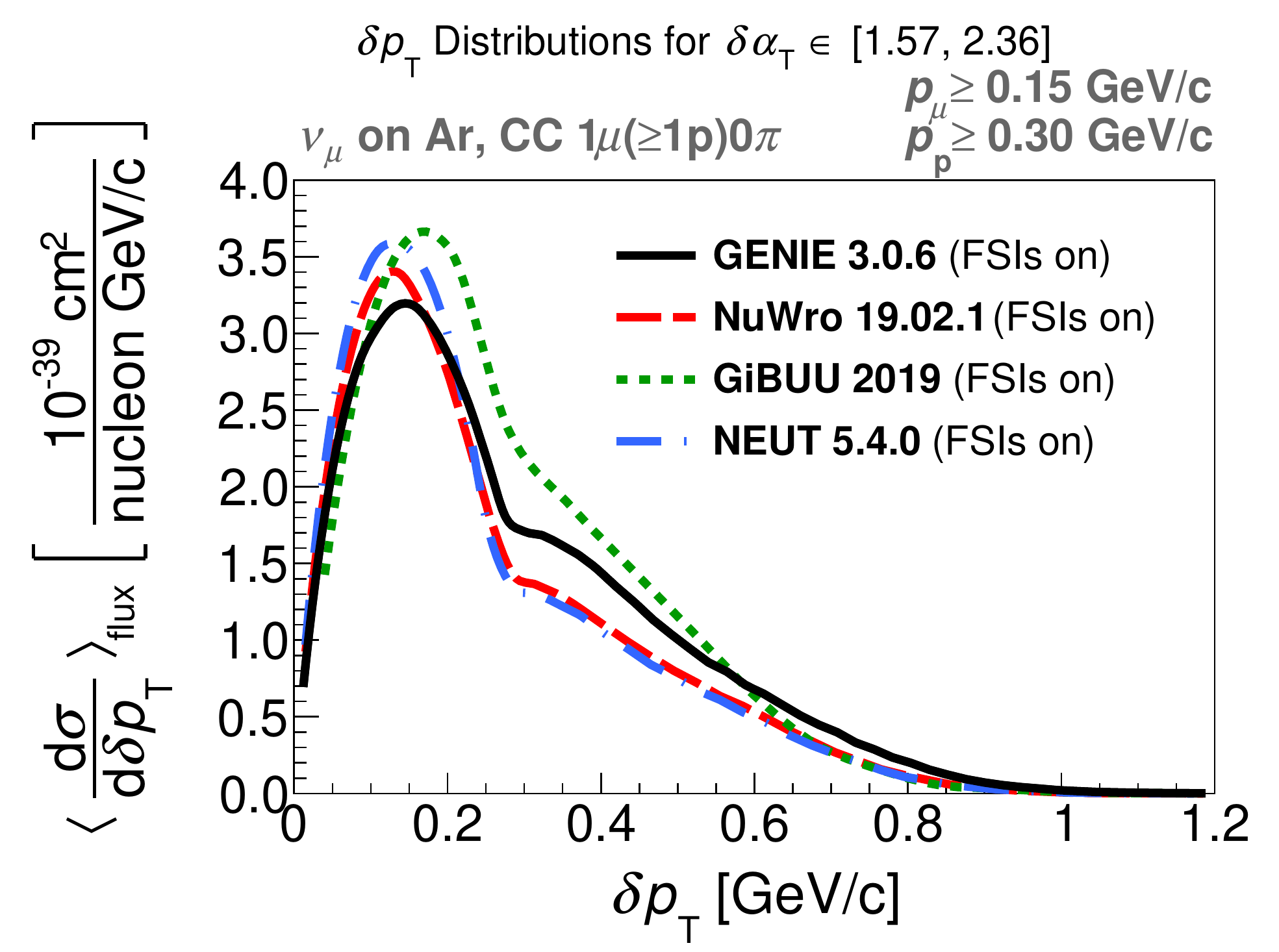}
		\vspace{-6.mm}
		\caption{}
		\label{fsi_gencomp_dpt_dalphat_3}
	\end{subfigure}
\end{figure}
\begin{figure}
\ContinuedFloat
	\begin{subfigure}[b]{.48\textwidth}
			\includegraphics[width=\textwidth]{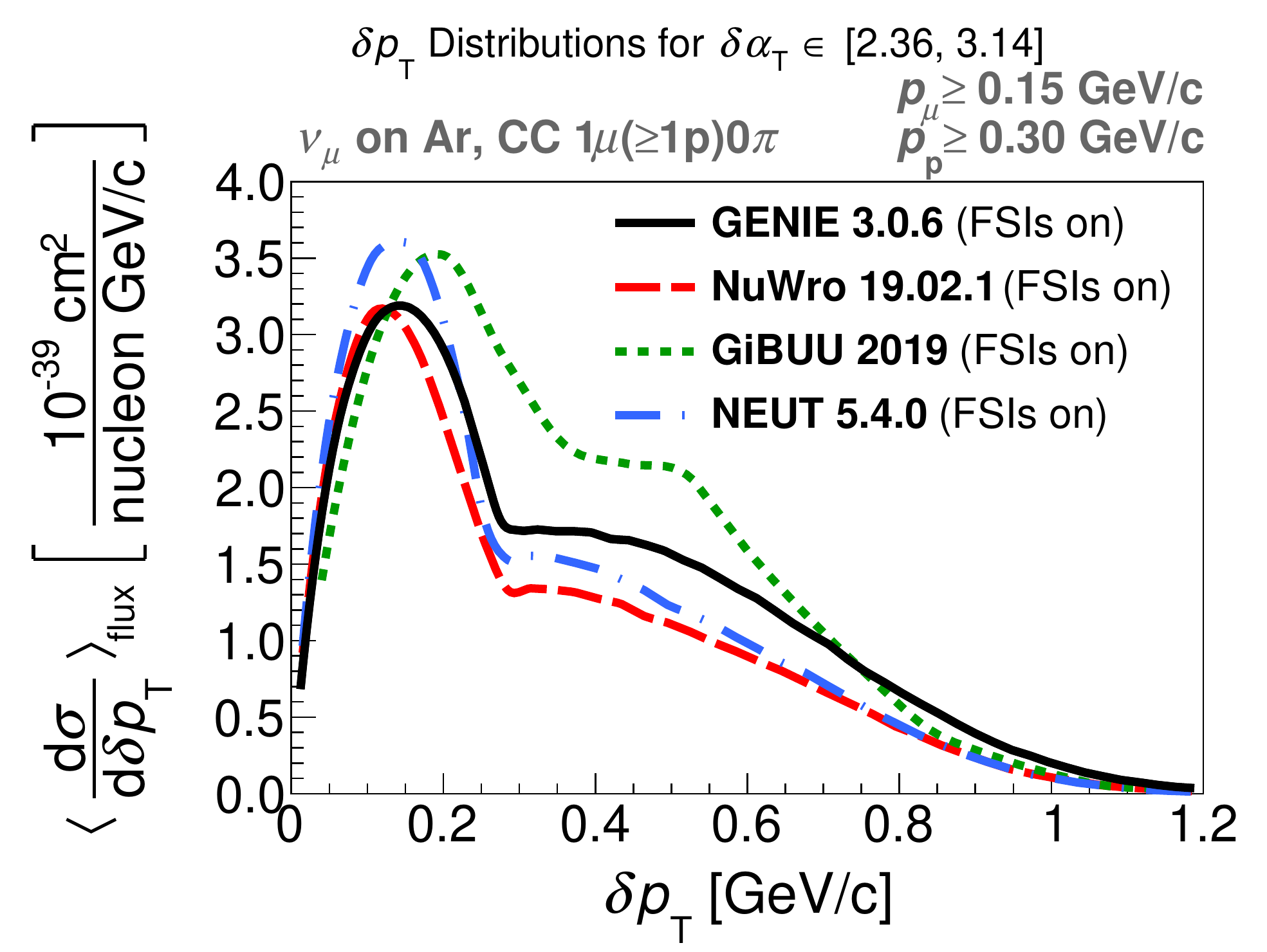}
			\vspace{-6.mm}
			\caption{}
			\label{fsi_gencomp_dpt_dalphat_4}
		\end{subfigure}
	\caption{Differential cross section of muon neutrino CC interactions on argon for the four event generators as a function of $\delta p_\text{T}$ for four different ranges of $\delta\alpha_\text{T}$. The range of  $\delta\alpha_\text{T}$ varies from the lowest values (\subref{fsi_gencomp_dpt_dalphat_1} to the highest ones (\subref{fsi_gencomp_dpt_dalphat_4}). For all selected events the leading proton momentum is at least $\SI{0.300}{\GeV/\text{c}}$ and the muon momentum is at least $\SI{0.150}{\GeV/\text{c}}$.}
	\label{fsi_gencomp_dpt_dalphat_slices}
\end{figure}
The generator predictions for individual bins are shown in order of increasing $\delta\alpha_\text{T}$  as separate subfigures. The degree of separation between the GiBUU result (dotted green) and those from the other generators is comparable at low and intermediate $\delta\alpha_\text{T}$ values to that already seen in Fig.~\ref{fsi_gencomp_dpt} for the single-differential $d\sigma/d\delta p_\text{T}$ cross section. However, in the highest $\delta\alpha_\text{T}$ bin shown in Fig.~\ref{fsi_gencomp_dpt_dalphat_4}, GiBUU's large $n$p-$n$h component drives its prediction away from the others, both via a shift in the low-$\delta p_\text{T}$ peak and an enhancement to the tail at moderate $\delta p_\text{T}$. This difference between the generator models may be worthy of future experimental investigation by MicroBooNE or another LArTPC detector in the SBN program.

\subsection{TKI in two-proton events}
Although it is known to play an important role in neutrino-nucleus interactions at accelerator energies, the 2p-2h contribution to the cross section is still poorly understood, and model predictions available in generators vary widely~\cite{SuSAv2GENIE}. Future detailed measurements of two-proton neutrino scattering events by LArTPCs, building upon pioneering work by the ArgoNeuT collaboration~\cite{ArgoNeuT2p}, may shed important light on this topic.

\begin{figure}[htbp!]
	\begin{subfigure}[b]{.48\textwidth}
		\includegraphics[width=\textwidth]{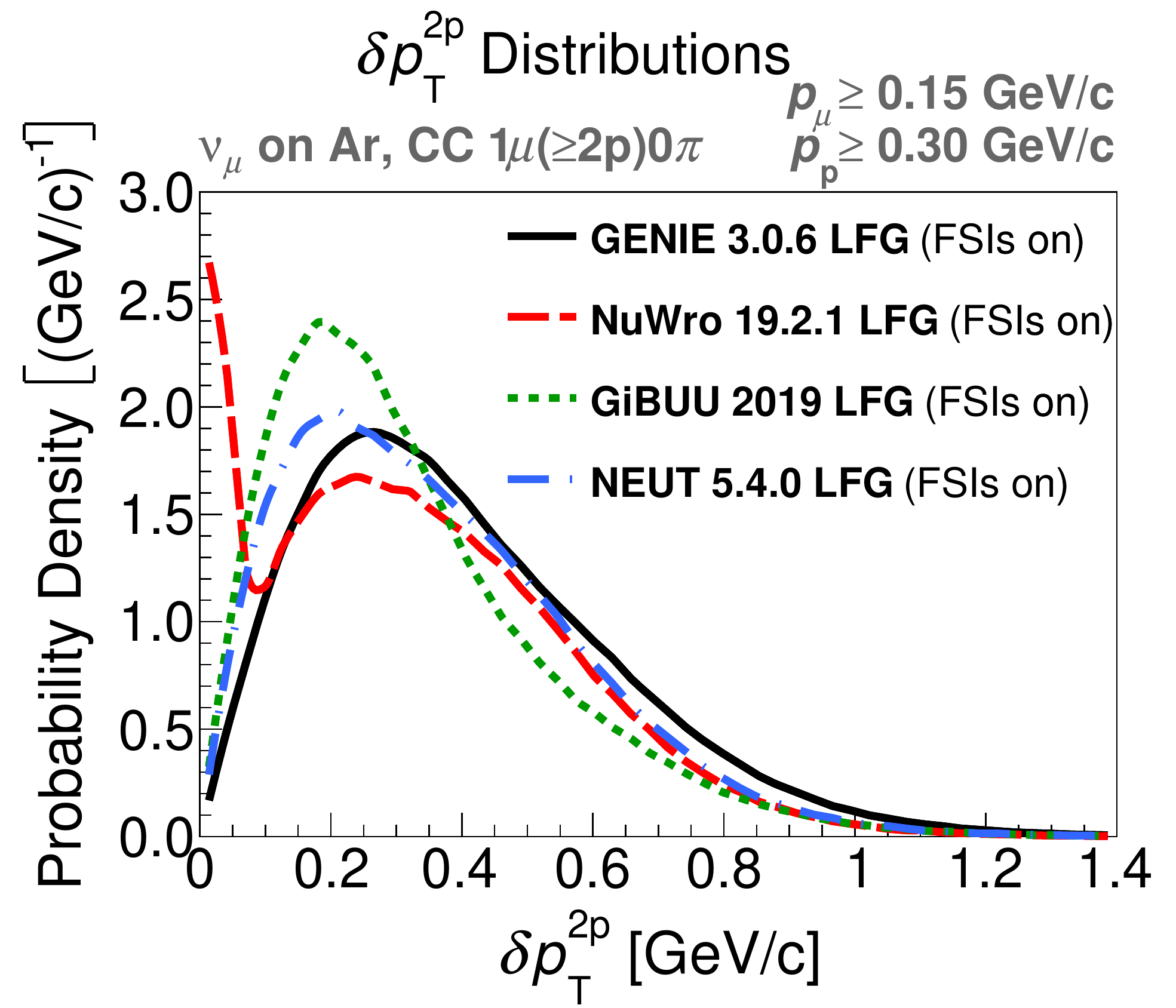}
		\vspace{-5.mm}
		\caption{}
		\label{new_stvs_dpt2p}
	\end{subfigure}
	\begin{subfigure}[b]{.48\textwidth}
		\includegraphics[width=\textwidth]{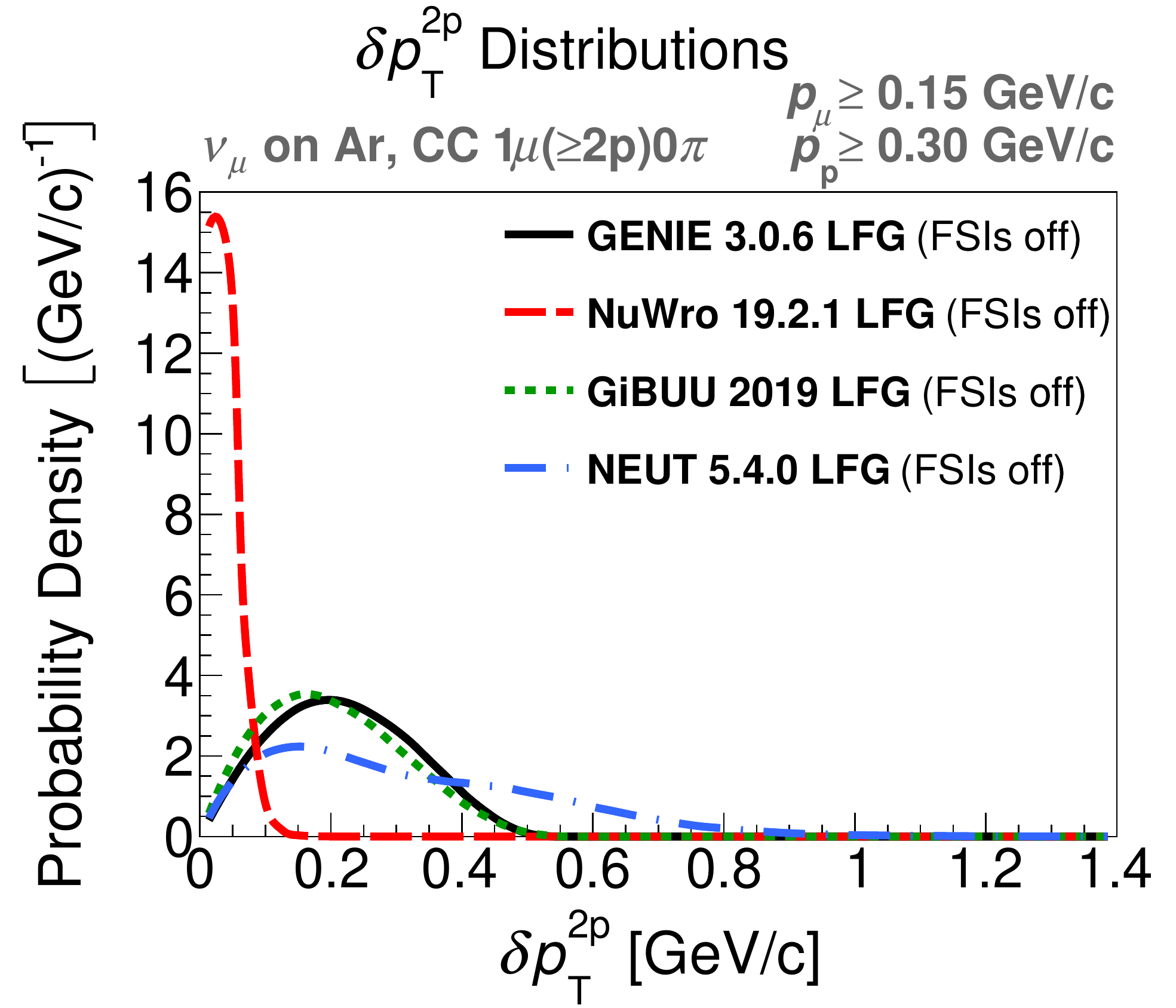}
		\vspace{-5.mm}
		\caption{}
		\label{new_stvs_fsi_off_2}
	\end{subfigure}
	\caption{Predictions of the four generators for an additional TKI variable distribution of $\delta p_\text{T}^{\hspace{.2ex} 2\text{p}}$ with FSIs turned on \subref{new_stvs_dpt2p} and turned off \subref{new_stvs_fsi_off_2}. For all selected events, the momentum of each of the two leading protons is at least $\SI{0.300}{\GeV/\text{c}}$, and the muon momentum is at least $\SI{0.150}{\GeV/\text{c}}$.}
	\label{new_stvs}
\end{figure}

As discussed in Sec.~\ref{stki}, an additional TKI variable unique to two-proton final states, $\delta \vec{p}_\text{T}^{\hspace{.2ex} 2\text{p}}$, can be calculated according to Eq.~\ref{dpt_2p}. Distributions of the magnitude $\delta p_\text{T}^{\hspace{.2ex} 2\text{p}}$ of this variable as predicted by the four event generators under study are shown in Fig.~\ref{new_stvs}. The signal definition is nearly the same as the one used previously, with the sole additional requirement being that a second final-state proton must have a momentum greater than or equal to \SI{0.3}{\GeV/\text{c}}. Fig.~\ref{new_stvs_dpt2p} compares the $\delta p_\text{T}^{\hspace{.2ex} 2\text{p}}$ distributions using the full interaction simulation available in each generator, while Fig.~\ref{new_stvs_fsi_off_2} does the same while neglecting hadronic FSIs. To emphasize differences in the shapes of the distributions, each differential cross section has been renormalized to integrate to unity over the region shown. The y-axis in both subfigures therefore indicates that the plotted quantity is a probability density function for $\delta p_\text{T}^{\hspace{.2ex} 2\text{p}}$.

A striking difference between the NuWro result and the other generator predictions is present in both plots at low values of $\delta p_\text{T}^{\hspace{.2ex} 2\text{p}}$. The large peak near zero uniquely predicted by NuWro is a consequence of different modeling assumptions about the initial-state nucleon pair in 2p-2h events. While the initial pair is assembled by the other generators by independently sampling two nucleon momenta from the ground-state nuclear model, NuWro typically assumes that the scattering occurs on a back-to-back nucleon pair~\cite{NuWroBackToBack}. In such events the transverse momenta of the muon and of the outgoing nucleons are nearly in balance, which leads to more events with low $\delta p_\text{T}^{\hspace{.2ex} 2\text{p}}$. This interpretation is supported by examining the distributions in Fig.~\ref{new_stvs_fsi_off_2}, which were calculated while neglecting FSIs. With FSIs removed, only the low $\delta p_\text{T}^{\hspace{.2ex} 2\text{p}}$ peak remains. This handling of the initial state in 2p-2h interactions is distinct to NuWro and might be explored via an experimental measurement of $\delta p_\text{T}^{\hspace{.2ex} 2\text{p}}$ with MicroBooNE or another LArTPC detector installed in a neutrino beam. While combining measurements of both final-state proton momenta to calculate $\delta p_\text{T}^{\hspace{.2ex} 2\text{p}}$ could entail some experimental challenges due to limited resolution (particularly for protons near threshold), the peak near zero in Fig.~\ref{new_stvs_dpt2p} is distinct enough that its presence or absence in the data might still be reliably verified.

\section{Summary and conclusions}
The complexity of neutrino interactions arising from nuclear effects is an important topic and significant efforts to predict them using intertwined interaction models implemented in event generators have been made. Precise modeling of neutrino-argon scattering demands a better understanding of nuclear effects and is of interest for future neutrino oscillation programs such as SBN and DUNE. Transverse Kinematic Imbalance has been shown to be a powerful tool to investigate these nuclear effects in experimental studies of neutrino scattering on hydrocarbon, but a comparable measurement for an argon target is yet to emerge in the scientific literature. The generator comparison studies shown in this paper suggest that a future measurement of TKI observables from a LArTPC neutrino experiment will be valuable for improving our understanding of the relevant nuclear physics. Specific examples considered in detail include (1)~the relative importance of CCQE and CC2p-2h contributions in explaining the asymmetry of the $\delta\alpha_\text{T}$ differential cross section, and (2)~the configuration of the initial two-nucleon system in 2p-2h events. The generator comparisons considered here are highlights from a larger set of studies presented in Ref.~\cite{bathe-peters:2020}.

\begin{acknowledgments}
We thank X.-G. Lu for helpful comments and advice. R. Guenette and L. Bathe-Peters are partly supported by the Department of Energy grant award number DE-SC0007881. This manuscript has been authored by Fermi Research Alliance, LLC under Contract No. DE-AC02-07CH11359 with the U.S. Department of Energy, Office of Science, Office of High Energy Physics.
\end{acknowledgments}

\appendix
\section{Summary of neutrino-nucleus interaction model elements} \label{model_elements}
The neutrino-nucleus interaction models available in GENIE~$3.0.6$, NuWro~$19.01.1$, GiBUU~$2019$ and NEUT~$5.4.0$ are summarized in Table~\ref{tab:generators}, and the models used to obtain the results shown in this paper are highlighted in green.
	\begin{table}[H]
		\caption{Summary of neutrino-nucleus interaction model elements available in GENIE, NuWro, GiBUU and NEUT. The models used for the results presented in this paper in Sec.~\ref{results} are highlighted in green.}
		\resizebox{\columnwidth}{!}{%
			\begin{tabular}{|c||*{5}{c|}}
				\hline
				\diagbox{\textbf{Model element}}{\textbf{Generator}} & $\vcenter{\hbox{\includegraphics[width=3cm]{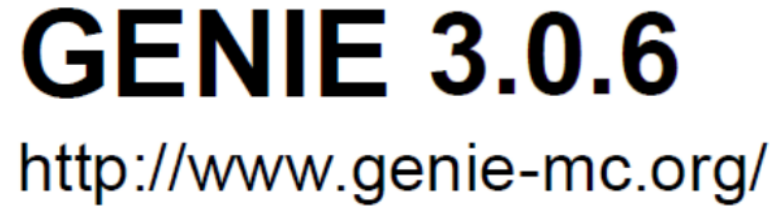}}}$ \hspace{2mm} $\vcenter{\hbox{\includegraphics[height=1.2cm]{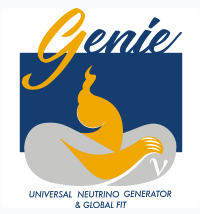}}}$ &
					$\vcenter{\hbox{\includegraphics[height=1.2cm]{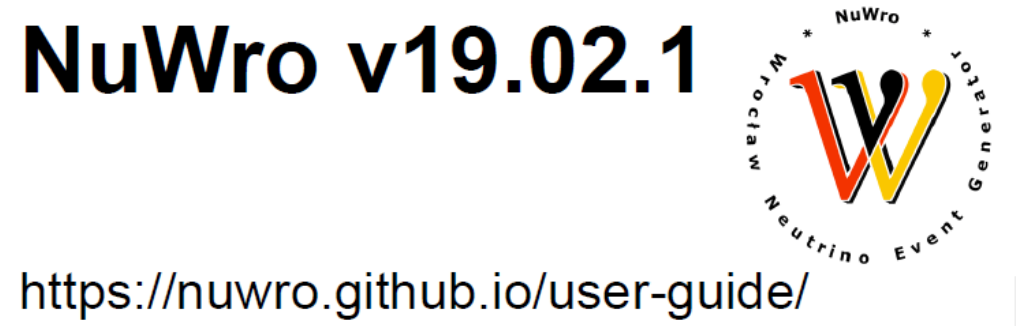}}}$ & $\vcenter{\hbox{\includegraphics[height=1.2cm]{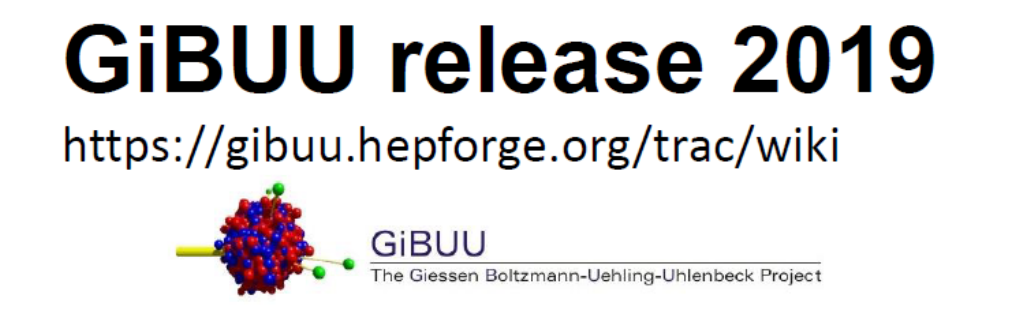}}}$ & $\vcenter{\hbox{\includegraphics[height=1.2cm]{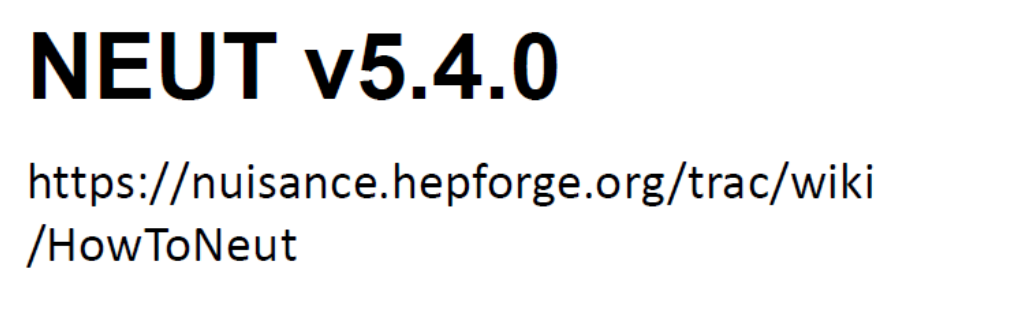}}}$ \\ \hline \hline
				\multirow{2}{*}{Nuclear Model} & Bodek-Ritchie RFG, {\color{table_color}\textbf{\textit{LFG}}} & Bodek-Ritchie RFG, & {\color{table_color}\textbf{\textit{LFG}}} (with local & Bodek-Ritchie RFG \\
				& SF, Effective SF & {\color{table_color}\textbf{\textit{LFG}}}, GFG, SF & dependent potential) & {\color{table_color}\textbf{\textit{LFG}}}, SF \\ \hline
				\multirow{2}{*}{Major CC Reaction Modes} & {\color{table_color}\textbf{\textit{QE}}}, {\color{table_color}\textbf{\textit{$n$p-$n$h}}}, {\color{table_color}\textbf{\textit{RES}}}, & {\color{table_color}\textbf{\textit{QE}}}, {\color{table_color}\textbf{\textit{$n$p-$n$h}}}, {\color{table_color}\textbf{\textit{RES}}}, & {\color{table_color}\textbf{\textit{QE}}}, {\color{table_color}\textbf{\textit{$n$p-$n$h}}}, {\color{table_color}\textbf{\textit{RES}}}, & {\color{table_color}\textbf{\textit{QE}}}, {\color{table_color}\textbf{\textit{$n$p-$n$h}}}, {\color{table_color}\textbf{\textit{RES}}}, \\
				& {\color{table_color}\textbf{\textit{DIS}}}, COH & {\color{table_color}\textbf{\textit{DIS}}}, COH & {\color{table_color}\textbf{\textit{DIS}}}, no COH & {\color{table_color}\textbf{\textit{DIS}}}, COH \\ \hline
				\multirow{2}{*}{CCQE Models} & Llewellyn-Smith, & {\color{table_color}\textbf{\textit{Llewelyn-Smith (+RPA)}}} & \multirow{2}{*}{{\color{table_color}\textbf{\textit{GiBUU QE model}}}} & Smith-Moniz (LS-like), \\
				& {\color{table_color}\textbf{\textit{Nieves}}} & SF, Effective Potential & & {\color{table_color}\textbf{\textit{Nieves}}} \\ \hline
				\multirow{2}{*}{QE Vector From Factor Model} & Dipole, BBBA$03$, & Dipole, BBBA$03$, & BBBA$03$, & Dipole, \\
				& {\color{table_color}\textbf{\textit{BBBA$\boldsymbol{05}$}}}, BBBA$07$ & {\color{table_color}\textbf{\textit{BBBA$\boldsymbol{05}$}}}, JLab, NN10 & {\color{table_color}\textbf{\textit{BBBA$\boldsymbol{05}$}}}, BBBA$07$ & {\color{table_color}\textbf{\textit{BBBA$\boldsymbol{05}$}}}, BBBA$07$ \\ \hline
				\multirow{2}{*}{QE Axial From Factor Model} & {\color{table_color}\textbf{\textit{Dipole}}}, Z-Expansion, & {\color{table_color}\textbf{\textit{Dipole}}}, 2-, 3-, 4-fold & \multirow{2}{*}{{\color{table_color}\textbf{\textit{Dipole}}}} & \multirow{2}{*}{{\color{table_color}\textbf{\textit{Dipole}}}} \\
				& Running MA & parabolic modification &  & \\ \hline
				\multirow{2}{*}{RPA Corrections for CCQE} & Built into {\color{table_color}\textbf{\textit{Nieves CCQE}}}, & No RPA, & \multirow{2}{*}{{\color{table_color}\textbf{\textit{Nieves-like}}}} & \multirow{2}{*}{{\color{table_color}\textbf{\textit{Nieves-like}}}} \\
				& None for LS CCQE & {\color{table_color}\textbf{\textit{RPA on}}} (not Nieves-like) & & \\ \hline
				\multirow{2}{*}{CC $n$p-$n$h Models} & {\color{table_color}\textbf{\textit{Nieves}}}, & TEM (Arie Bodek), & {\color{table_color}\textbf{Christy Model}} & \multirow{2}{*}{{\color{table_color}\textbf{\textit{Nieves}}}} \\
				& Empirical, Marteau, SuSAv2 & {\color{table_color}\textbf{\textit{Nieves}}}, SuSAv2 & for electrons & \\ \hline
				\multirow{2}{*}{CCRES Models} & {\color{table_color}\textbf{\textit{Rein-Sehgal}}}, & {\color{table_color}\textbf{\textit{Adler-Rarita}}} & \multirow{2}{*}{{\color{table_color}\textbf{\textit{GiBUU RES model}}}} & {\color{table_color}\textbf{\textit{Rein-Sehgal}}}, \\
				& Berger-Sehgal & {\color{table_color}\textbf{\textit{Schwinger}}} & & Berger-Sehgal \\ \hline
				\multirow{2}{*}{CCDIS Models} & \multirow{2}{*}{{\color{table_color}\textbf{\textit{Bodek-Yang}}}} & \multirow{2}{*}{{\color{table_color}\textbf{\textit{Bodek-Yang}}}} & \multirow{2}{*}{{\color{table_color}\textbf{\textit{data-driven}}}}
				& \multirow{2}{*}{{\color{table_color}\textbf{\textit{Bodek-Yang}}}} \\
				&  &  & & \\ \hline
				\multirow{2}{*}{COH Production} & {\color{table_color}\textbf{\textit{Rein-Sehgal}}}, & {\color{table_color}\textbf{\textit{Rein-Sehgal}}}, & \multirow{2}{*}{not available} & \multirow{2}{*}{{\color{table_color}\textbf{\textit{Rein-Sehgal}}}} \\
				& Berger-Sehgal & Berger-Sehgal &  & \\ \hline
				\multirow{2}{*}{FSI} & INTRANUKE hA, & {\color{table_color}\textbf{\textit{Metropolis and Oset cascade}}} & {\color{table_color}\textbf{\textit{BUU Quantum-}}} & {\color{table_color}\textbf{\textit{Semi-classical cascade}}} \\
				& {\color{table_color}\textbf{\textit{hN}}} & (hN-like) & {\color{table_color}\textbf{\textit{Kinetic Transport}}} & (hN-like) \\ \hline
			\end{tabular}
		}
		\label{tab:generators}
	\end{table}
A summary of the most important parameter and model element settings for all simulation tools used for this study can be found in Table~\ref{tab:settings}.
\begin{table}[h!]
\begin{flushright}
	\caption[Model Choices and Parameter Settings for Neutrino Event Simulations]{Model choices and parameter settings for the neutrino event simulations prepared for this paper. If one setting in this table is not available in GENIE, NuWro, GiBUU and NEUT, the closest model available was used for the neutrino event simulation. The choices are highlighted in green in Table~\ref{tab:generators}.}
			\resizebox{\columnwidth}{!}{%
	\begin{tabular}{|l||c|}
		\hline
		\textbf{Model element}	& \textbf{Setting} \\ \hline \hline
 Interaction Types             	& QE, $n$p-$n$h, RES, DIS, COH (CC only) \\ \hline
 Nuclear Model            	    & Local Fermi Gas \\ \hline
 Nuclear Potential 	            & Woods-Saxon Shape \\ \hline
 Coulomb Potential for Nucleons & Disabled \\ \hline
 CCQE Model  	        & Nieves \\ \hline
 QE Vector Form Factors           		& BBA$05$ \\ \hline
 QE Axial Form Factor	    		    & Dipole \\ \hline
  RPA Corrections for CCQE		& Enabled \\ \hline
 CCQE Effective Mass Corrections     & Disabled \\ \hline
 CC$n$p-$n$h Model              		& Nieves, Christy model for GiBUU \\ \hline
 CCRES model        	    	& Rein-Sehgal \\ \hline
 Final-State Interactions (FSIs)	                    	& Generator-specific implementations \\ \hline
 Pauli-Blocking                	& Enabled \\ \hline
 CCQE Axial Mass		        & $M_\text{A} = 1.0$ GeV \\ \hline
 Pion Axial Mass		        & $M_\pi = \SI{1.0}{\GeV}$ \\ \hline
 Nucleon Binding Energy            & $E_\text{b} = \SI{30}{\MeV}$ \\ \hline
 Fermi Momentum                 & $p_\text{F} = \SI{0.220}{\GeV/\text{c}}$ \\ \hline
	\end{tabular}
	}
	\label{tab:settings}
	\end{flushright}
\end{table}
\subsection{How to reproduce the results with GENIE}
The GENIE model set used in this paper is based on the $\mathtt{G18\_10b\_00\_000}$ configuration. The following minor changes were made for the studies described herein:
\begin{itemize}
    \item The nucleon binding energy in \isotope[40]{Ar} was changed from 29.5~MeV to 30~MeV.
    \item The CCQE axial mass was changed from 0.99~GeV to 1.00~GeV.
    \item The parameterization for the CCQE vector form factors was changed from BBA07 to BBA05~\cite{Bradford:2006yz}.
    \item The RES model (but not the COH model) was changed to one based on the Rein-Sehgal calculation~\cite{Rein:1980wg}.
    \item The RES axial mass was changed from 1.12~GeV to 1.00~GeV.
\end{itemize}
From an unaltered installation of GENIE~3.0.6, one may obtain a model configuration identical to ours by copying the three XML files included in the supplemental materials (\texttt{CommonParam.xml}, \texttt{ModelConfiguration.xml}, and \texttt{ReinSehgalRESPXSec.xml}) into the \texttt{config/G18\_10b/} subfolder of the GENIE source code. The \texttt{ModelConfiguration.xml} file should replace the existing one in that subfolder, while the other two XML files are new additions. After this is done, the total cross-section spline files needed to run the simulation should be regenerated in the usual way using the \texttt{gmkspl} command-line tool with the option \texttt{--tune G18\_10b\_00\_000}.
\subsection{How to reproduce the results with NuWro}
The MicroBooNE flux file (see ancillary $\mathtt{flux.txt}$ file) has to be formatted into the NuWro beam $\mathtt{.txt}$-file format. Then, the beam type, particle and direction has to be specified to (in the case of MicroBooNE with a single flavor beam):
\begin{equation}
    \begin{aligned}
        &\verb!beam_type=0! \\
        &\verb!beam_particle= 14! \\
        &\verb!beam_direction = 0 0 1!
    \end{aligned}
\end{equation}
Moreover the energy range on the $x$-axis of the flux file has to be specified as the first two entries at $\mathtt{beam\_energy}$. This should be given in MeV. The example input called $\mathtt{params.txt}$ \footnote{\url{https://github.com/NuWro/nuwro/tree/master/data}} was modified in the following way: The parameters for a beam defined by hand are commented out and the generated MicroBooNE flux file is specified. The nuclear target is set to argon. All CC dynamics channels for event generation are enabled, and all NC channels are disabled. The QE electromagnetic form factors are chosen to follow the BBBA05 parametrisation in case of the QE vector form factor, and the QE axial form factor is described by a dipole. The CCQE and CCRES axial masses are both adjusted to \SI{1000}{MeV}. The coherent scattering model is changed from the Berger-Sehgal to the Rein-Sehgal model. The $n$p-$n$h model used is the Nieves model. The nucleon binding energy is set to \SI{30}{\MeV} and the Fermi momentum remains unchanged with $p_F = \SI{220}{\MeV}$. The target nucleus description is performed with the Local Fermi Gas model (default setting). FSIs are turned on or off according to the specific plot by setting $\mathtt{kaskada\_on}$ to $1$ or $0$ respectively in the NuWro input $\mathtt{params.txt}$-file. There are no other changes to be made in order to reproduce the NuWro results presented in this paper. For more specifics, see the ancillary $\mathtt{params\_LFG\_fsi\_off.txt}$ file.
\subsection{How to reproduce the results with GiBUU}
After installing the GiBUU $2019$ release, the \texttt{testRun/jobCards/005\_Neutrino\_MicroBooNE-nu.job} file is modified in the following ways. The MicroBooNE flux file (see the ancillary \texttt{flux.txt} file) has to be formatted into the GiBUU beam $\mathtt{.dat}$-file format. It should then be copied into into the subfolder $\mathtt{buuinput2019/neutrino/}$. The source file
\texttt{/release2019/code/init/neutrino\allowbreak/expNeutrinofluxes.f90} contains the Fortran code which handles the experimental flux specifications.
After setting all subprocesses to true, the Fermi momentum is adjusted from its default value to $p_F = \SI{220}{\MeV}$. The number of events to be generated is proportional to
\begin{equation} \label{gibuu_product}
    \verb!target_A*numEnsembles*num_runs_SameEnergy!
\end{equation}
and can be varied by adjusting any of the values in the product (\ref{gibuu_product}). We set these parameters to $\verb!target_A! = 40$, $\verb!num_Ensembles! = 4000$ and $\verb!num_runs_SameEnergy! = 70$. The latter also sets the number of GiBUU output $\verb!.root!$-files produced. FSIs can be turned off by setting the number of time steps in the numerical integration to zero. That is, one may disable FSIs in the job card file via the parameter setting
\begin{equation} \label{gibuu_fsi}
    \verb!numTimeSteps=0!
\end{equation}
The models available in GiBUU are chosen by commenting and uncommenting respective lines in the job card file. The axial mass form factor for resonance production should be set to $\SI{1.00}{\GeV}$. We also enabled RPA corrections by adding the following lines to the job card file:
\begin{equation}
    \begin{aligned}
        &! \verb! file: code/init/lepton/matrixElementQE.f90! \\
        &\verb!&MatrixElementQE! \\
        &\verb!    useCorrelations=.true.! \\
        &\verb!    nievesCorr_para=2! \\
        &\verb!\!
    \end{aligned}
\end{equation}
The RES model is switched to the Rein-Sehgal model and the QE form factor parametrization is adjusted to BBBA$05$. The parameters $\verb!useNonStandardMA!$ and $\verb!mesonMesonScattering!$ are set to true. There are no other changes to be made in order to reproduce the GiBUU results presented in this paper. For more specifics, see the ancillary $\mathtt{005\_neutrino\_MicroBooNE\_numu\_GiBUU\_fsi\_off.job}$ file.
\subsection{How to reproduce the results with NEUT}
After downloading and installing NEUT $5.4.0$, example NEUT cards can be found in the $\mathtt{src/neutsmpl/Cards/}$ directory. The following changes to the NEUT card $\mathtt{neut\_5.4.0\_nd5\_Ar.card}$ have to be made in order to reproduce the results presented in this paper. The number of events is set to $5000000$. The ancillary $\mathtt{flux.txt}$ file can be formatted into a $\mathtt{.root}$ file and specified in the NEUT card instead of the default flux file. This $\mathtt{.root}$ file has to contain a histogram with the MicroBooNE flux that also has to be specified. The nucleon binding energy $\mathtt{NEUT-EBIND}$ ("Nucleon\_Binding\_Energy\_Ground\_State = Total Binding energy - Excitation energy (\isotope[40]{Ar}) $= \SI{30}{\MeV} - \SI{9.9}{\MeV} = \SI{20.1}{\MeV}$") and $\mathtt{NEUT-EBINDCORR}$ is set to $1$. The scaling factors applied to various interaction modes are set to $1$ for CCQE, CC$1\pi$, CC DIS $1320$, COH, CC $\eta$, CC K, CCDIS, CC $1\gamma$, CC$2$p-$2$h and CC diffractive interactions. All other scaling factors are set to $0$. The $1$p-$1$h Nieves model is combined with a Smith-Moniz model, Local Fermi Gas model and the BBBA$05$ form factor parametrization. Both the QE axial mass and the pion axial mass $\mathtt{NEUT-XMARSRES}$ are set to $\SI{1}{\GeV}$. Further changes include additional lines to turn on Fermi motion and Pauli-blocking. The coherent pion production model is also set to be the Rein-Sehgal calculation. FSIs can be turned on or off with the $\mathtt{NEUT-RESCAT}$ parameter. For more specifics, see the ancillary $\mathtt{MicroBooNE\_numu\_LFG\_fsi\_off\_neut.card}$ file. \\ \newpage

\section{TKI topology for two-proton events}
\begin{figure}[hbt!]
	\centering
	\includegraphics[width=.49\textwidth]{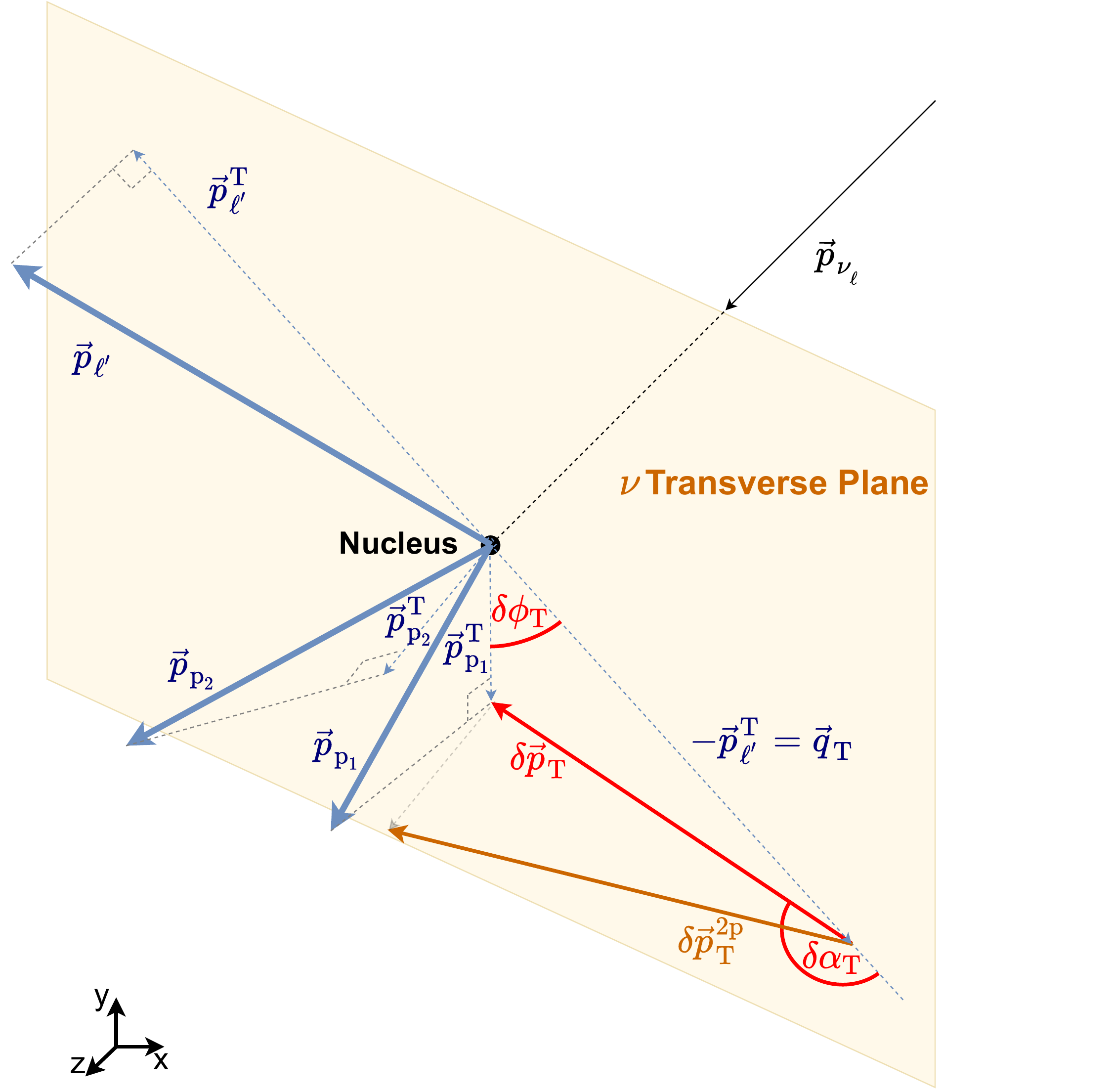}
	\caption[Illustration for the Construction of Additional TKI variables]{Illustration for the construction of the additional TKI variable $\delta \vec{p}_\text{T}^{\hspace{.2ex} 2\text{p}}$ (see Eq.~\ref{dpt_2p}). In addition to the transverse momentum imbalance $\delta \vec{p}_\text{T}$ (red) illustrated in Fig.~\ref{stki} that considers the leading proton (p$_1$), the new TKI variable $\delta\vec{p}_\text{T}^{\hspace{.2ex} 2\text{p}}$ (dark yellow) includes the next-to-leading proton (p$_2$) for the calculation of the transverse momentum imbalance.}
	\label{STKI_add}
\end{figure}

To aid the reader in visualizing the construction of the additional TKI variable $\delta \vec{p}_\text{T}^{\hspace{.2ex} 2\text{p}}$, a supplemental illustration is provided in Fig.~\ref{STKI_add}.
%
\bibliography{bibliography}
\end{document}